\documentclass{acmsmall} 
\usepackage{amsmath,amssymb}
\usepackage{epsfig}
\usepackage{graphicx}
\usepackage{subfigure}
\usepackage{epstopdf}
\usepackage{ifthen}
\usepackage{color}
\usepackage{keyval}
\usepackage[colorlinks=true,linkcolor=blue,citecolor=blue]{hyperref}
\usepackage[ruled]{algorithm2e}
\usepackage{cases}

\usepackage{thmtools}
\usepackage{thm-restate}

\setcopyright{none}
\SetAlFnt{\small}
\SetAlCapFnt{\small}
\SetAlCapNameFnt{\small}
\SetAlCapHSkip{0pt}
\IncMargin{-\parindent}

\newtheorem{fact}{fact}
\newcommand{\para}[1]{{\vspace{3pt} \bf \noindent #1 \hspace{5pt}}}

\newcommand{\chgins}[1]{\textcolor{blue}{#1}}
\DeclareMathOperator*{\argmax}{argmax}

\begin{document}

\markboth{Zhi Yang and Wei Chen.}{A Game Theoretic Model for the Formation of
Navigable Small-World Networks
}

\title{ 
A Game Theoretic Model for the Formation of \\
Navigable Small-World Networks --- the Tradeoff between Distance and Reciprocity}
\author{ZHI YANG
\affil{Peking University}
WEI CHEN
\affil{Microsoft Research}
}

\begin{abstract}
Kleinberg proposed a family of small-world networks to explain the
navigability of large-scale real-world social networks. However, the
underlying mechanism that drives real networks to be navigable is
not yet well understood. In this paper, we present a game theoretic
model for the formation of navigable small world networks. We model
the network formation as a game called the {\em Distance-Reciprocity Balanced (DRB)} game
	in which people seek for both high
reciprocity and long-distance relationships. We show that the game has only
	two Nash equilibria: One is the navigable small-world network, and
	the other is the random network in which each node connects
	with each other node with equal probability, and any other network state
	can reach the navigable small world via a sequence of best-response moves of nodes.
We further show that the navigable small world equilibrium is very stable ---
	(a) no collusion of any size would benefit from deviating from it; and
	(b) after an arbitrary deviations of a large random set of nodes,
	the network would return to the navigable small world as soon as every node
	takes one best-response step.
In contrast, for the random network, a small group collusion or random perturbations is guaranteed to
	bring the network out of the random-network equilibrium and move to the navigable network as soon as every node takes one
	best-response step.
Moreover, we show that navigable small world equilibrium has much better social welfare than the random network,
	and provide the price-of-anarchy and price-of-stability results of the game.
Our empirical evaluation further demonstrates that the system always converges to
the navigable network even when limited or no information about
other players' strategies is available, and the DRB game simulated on real-world networks
	leads to navigability characteristic that is very close to that of the real networks, even though
	the real-world networks have non-uniform population distributions different from the
	Kleinberg's small-world model.
Our theoretical and
empirical analyses provide important new insight on the connection
between distance, reciprocity and navigability in social networks.
\end{abstract}



\begin{bottomstuff}
A preliminary version of this work appeared in Proceedings of the 24th International World Wide Web Conference (WWW 2015).
The current version contains a number of new results comparing to the preliminary version, such as heterogeneous utility functions,
	analytical results on the best-response dynamics, non-existence of non-uniform equilibria, price of
	anarchy and price of stability, and empirical evaluation on real datasets.

The work was partly done when the
first author was a visiting researcher at Microsoft Research
Asia.
This work was supported by the National Basic Research
Program of China (Grant No. 2014CB340400).

Author's addresses: Zhi Yang, Computer Science Department, Peking University, Beijing, China; email:yangzhi@pku.edu.cn.
Wei Chen, Microsoft Research, Beijing, China; email:weic@microsoft.com.
\end{bottomstuff}

\maketitle

\section{Introduction}
In 1967, Milgram published his work on the now famous small-world
    experiment~\cite{Milgram1967small}: he asked test subjects to forward a letter
    to their friends in order for the letter to reach a person not known
    to the initiator of the letter.
He found that on average it took only six hops to connect two people in
    U.S., which is often attributed as the source of the popular term
    {\em six-degree of separation}.
This seminal work inspired numerous studies on the small-world phenomenon and
    small-world models, which last till the present day of information age.

In~\cite{Watts1998collective} Watts and Strogatz investigated a number of
    real-world networks such as film actor networks and power grids, and showed
    that many networks have both low diameter and high clustering (meaning
    two neighbors of a node are likely to be neighbors of each other), which
    is different from randomly wired networks.
They thus proposed a small-world model in which nodes are first placed on
    a ring or a grid with local connections, and then some connections
    are randomly
    rewired to connect to long-range contacts in the network.
The local and long-range connections can also be viewed as strong
ties and weak ties respectively in social relationships originally
proposed by
Granovetter~\cite{Granovetter1973strength,Granovetter1974getting}.

Kleinberg notices an important discrepancy between the small-world model
    of Watts and Strogatz and the original Milgram experiment:
    the latter shows not only that the average distance between nodes in
    the network are small, but also that a \emph{decentralized} routing algorithm using only local information can
construct short paths.  Here, we call a routing algorithm decentralized
in that given a source node $u$ and a destination node $v$, the algorithm attempts to come up with a path
$u = x_0, x_1, x_2, \ldots, x_m = v$, only using the acquaintance relationships of these $m$ intermediate nodes $x_0, x_1, x_2, \ldots, x_{m-1}$.
By contrast, a centralised algorithm (e.g., Dijkstra's algorithm) requires the nodes to know
full network (i.e., the acquaintance relationships among
all people in the world) to find an
optimal route, but obviously they cannot know this in real networks.

To address this issue, Kleinberg adjusted the Watts-Strogatz model so that
    the long-range connections are selected not uniformly at random among
    all nodes but inversely proportional to a power of the grid distance between the
    two end points of the connection~\cite{Kleinberg00}.
More specifically, Kleinberg modeled a social network as composed of $n^k$ nodes
    on a $k$-dimensional grid, with each node having local contacts to
    other nodes in its
    immediate geographic neighborhood.
Each node $u$ also establishes a number of long-range contacts, and
    a long-range link from $u$ to $v$ is established with probability
    proportional to $d_M(u,v)^{-r}$, where $d_M(u,v)$ is the grid distance
    between $u$ and $v$, and $r\ge 0$ is
    the model parameter indicating how likely nodes prefer to connect to
    remote nodes, which we call {\em connection preference} in the paper.
The Watts-Strogatz model corresponds to the case of $r=0$, and as $r$ increases,
    nodes are more likely to connect to other nodes in their vicinity.
Kleinberg modeled Milgram's experiment as
    decentralized greedy routing in such networks,
    in which each node
    only forwards messages to one of its neighbors with coordinate closest to
    the target node.
He showed that when $r=k$, greedy routing can be done efficiently in
    $O(\log^2 n)$ time in expectation, but for any $r\ne k$, it requires
    $\Omega(n^c)$ time for some constant $c$ depending on $r$, exponentially worse
    than the case of $r=k$.
Therefore, the small world at the critical value of $r=k$ is meant to model
    the real-world navigable network validated by Milgram and others' experiments,
    and we call it the {\em navigable small-world network}.

\begin{figure}[t]
\begin{minipage}[t]{0.49\columnwidth}
    \centerline{\includegraphics[width=0.8\textwidth, height =0.5\textwidth]{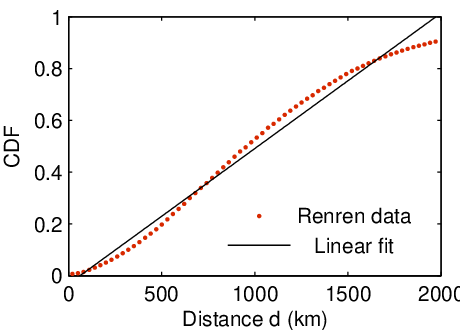}}
    \caption{The fraction of nodes within distance $d$ in Renren.}
    \label{fig:demension}
\end{minipage}
\hfill
\begin{minipage}[t]{0.49\columnwidth}
    \centerline{\includegraphics[width=0.8\textwidth, height =0.5\textwidth]{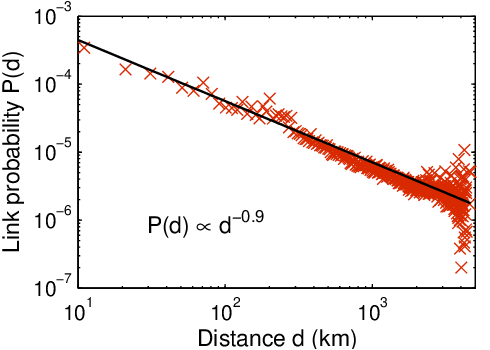}}
    \caption{Friendship probability vs. distance in Renren.}
    \label{fig:link_prob}
\end{minipage}
\end{figure}
After Kleinberg's theoretical analysis, a number of empirical studies have
    been conducted to verify if real networks indeed have connection
    preferences close to the critical value that allows efficient greedy
    routing \cite{Liben-Nowell,Adamic,Leskovec,Levy,Schaller}.
Since real population is not evenly distributed
    geographically as in the Kleinberg's model,
    Liben-Nowell et al.~\cite{Liben-Nowell}
    proposed to use the {\em fractional dimension} $D$,
    defined as the best value to fit
    $|\{w:d_M(u,w)\leq d_M(u,v)\}| = c\cdot d_M(u,v)^{D}$, averaged
over all $u$ and $v$.
They showed that when the connection preference
    $r=D$,
    the network is navigable.
They then studied a network of 495,836 LiveJournal users in the
    continental United States who list their hometowns, and find that
    $D \approx 0.8$ while $r=1.2$, reasonably close to $D$.
We apply the same approach to a ten million node
    Renren network~\cite{renren-imc10,Uncover-sybil},
    one of the largest online social networks in China. We map the hometown listed in users' profiles to (longitude, latitude)
    coordinates.
The resolution of our geographic data is
    limited to the level of towns and cities and thus we
    cannot get the exact distance of nodes within $10$km. We found that
    $D \approx 1$ (Figure~\ref{fig:demension})
    and $r \approx 0.9$
    (Figure~\ref{fig:link_prob}) in the Renren network.
Other studies~\cite{Adamic,Leskovec,Levy,Schaller} also reported
    connection preference $r$ to be close to $1$ in other online social networks
    (including Gowalla, Brightkite and Facebook).
Even though they did not report the fractional dimension, from both
    the LiveJournal data in~\cite{Liben-Nowell} and our Renren data, it
    is reasonable to believe that the fractional dimension is also close to $1$.
Therefore, empirical evidences all suggest that the real-world social networks
    indeed have connection preference close to the critical value and the
    network is navigable.

%
    A natural question to ask next is how navigable networks naturally emerge? What are the forces that make the connection preference
    become close to the critical value?
As Kleinberg pointed out in his survey paper~\cite{Kleinberg06} when talking
    about the above striking coincidence between theoretical prediction and
    empirical observation, ``it suggests that there may be deeper phenomena
    yet to be discovered here''. There are several studies trying to explain the emergence of navigable small-world networks
     \cite{Nisha01,hu,Aaron03,Sandberg06,haintreau08}, mostly by modeling certain underlying
     node or link dynamics (see additional related work below for more details).

In this paper, we tackle the problem in a novel way using a
game-theoretic approach, which is reasonable in
    modeling individual behaviors in social networks without central coordination.
One key insight we have is that connection preference $r$ is not a global
    preference but individual's own preference --- some prefer to connect
    to more faraway nodes while others prefer to connect to nearby nodes.
Therefore, we establish {\em small-world formation games} where individual
    node $u$'s strategy is its own connection preference $r_u$
    (Section~\ref{sec:game}).
This game formulation is different from most existing network formation games
    where individuals' strategies are creating actual links in the
    network (c.f.~\cite{AGT}). It allows us to directly explore the entire parameter space of connection preferences
    and answer the question on why nodes end up choosing a particular parameter setting leading to the navigable small world.

In terms of payoff functions, we first consider minimizing greedy routing distance
    to other nodes as the payoff, since it directly corresponds to
    the goal of navigable networks. However, Guly\'{a}s et al.~\cite{Andras} prove that
    with this payoff the navigable networks cannot emerge as a equilibrium for the one-dimensional case.
Our empirical analysis also indicates that nodes will converge to
    random networks ($r_u=0, \forall u$) rather than navigable
    networks for higher dimensions.
Our empirical analysis further shows that
    if we adjust the payoff with a cost proportional to the grid distance of
    remote connections, the equilibria are sensitive to the cost
    factor.

The above unsuccessful attempt suggests that besides the goal of shortening
    distance to remote nodes, some other natural objective may be in
    play. Reciprocity is regarded as a basic mechanism
    that creates stable social relationships in a person's life~\cite{Alvin}.
A number of prior works~\cite{Java,Liben-Nowell,Alan} also suggest
that people seek reciprocal relationships in online social networks.
Therefore, we propose a payoff function that is the product of
average distance
    of nodes to their long-range contacts and the probability of forming
    reciprocal relationship with long-range contacts.
We call this game the \emph{distance-reciprocity balanced (DRB)} game. In
practice, increasing relationship distance captures that individuals
attempt to create social bridges by linking to ``distant people'',
which can help them search for and obtain new resources. Meanwhile,
increasing reciprocity captures that individuals look at social
bonds by linking to ``people like them'', which could help them
preserve or maintain resources. Therefore, the DRB game is natural
since it captures sources of bridging and bonding social capital in
building social integration and solidarity \cite{captial}.
We further allow heterogeneous utility functions in that
	different users may weigh the tradeoff between distance and reciprocity in different ways.


Even though the payoff function for the DRB game is very simple, our analysis
    demonstrates that it is extremely effective in producing
    navigable small-world networks as the equilibrium structure.
In theoretical analysis (Section~\ref{sec:theory}),
    we first show that navigable small world
    ($r_u=k, \forall u$) and random small world ($r_u=0, \forall u$) are
    the only two Nash equilibria of the DRB game, despite the flexible and heterogeneous utility functions.
Moreover, for any strategy profile that is not the random network, it can always
	reach the navigable small world through a cascade of nearby nodes adopting strategy $k$
	in a best-response dynamic.

In terms of the stability of NE, we
prove that the navigable small world is a strong Nash equilibrium,
which means that it tolerates collusion of any size trying to gain
better payoff. Moreover, it also tolerates arbitrary deviations (without
the objective of increasing anyone's payoff) of large groups of
random deviators, since the system is guaranteed to return back to the navigable NE
	as soon as every node takes one best-response step.
In contrast, random small world can be moved away from its equilibrium state
	by either a random perturbation of one node or a collusion of two nearby nodes, and
	when a small random set of nodes perturb to different strategies, we prove that
	the system is guaranteed to converge to the navigable small world as soon as every node
	takes one best-response step.
Our theoretical analysis provides strong support
that the navigable small-world NE is the unique and stable equilibrium
that would naturally emerge in the DRB game.

We further examine the global function of social welfare (i.e., the total payoff
of all nodes) and how
	selfish behavior of users affect the social welfare.
Interestingly, we find that the global optimum can be reached by
	a fraction of nodes sacrificing their distance payoff to focus on reciprocity
	(by selecting a strategy greater than $k$)
	so that their neighbors could select strategy $k$ to reach a high balanced payoff of both
	distance and reciprocity.
This situation reminds us social relationships generated by different
	social status (e.g. employee-employer relationship) or by tight bonds with
	mutual understanding and support (such as marriages).
Next we compare the social welfare of navigable and random small-world networks with the
	global optimum through the standard price of anarchy (PoA) and price of stability (PoS)
	metrics, which is the ratio of social welfare between the global optimum and
	the worst (or the best) Nash equilibrium, respectively.
We show that navigable network has the better social welfare, and being only logarithmically worse than the global optimum.

To complement our theoretical analysis, we conduct empirical evaluations to cover more realistic
game scenarios
    not covered by our theoretical analysis
    (Section~\ref{sec:evaluation}).
We first test random perturbation cases and show that arbitrary initial
    profiles always converge to the navigable equilibrium in a few steps, while
    a very small random perturbation (less than theoretical prediction)
    of the random small world causes it to
    quickly converge back to the navigable equilibrium.
Next, we simulate more realistic scenarios where nodes have limited or no
    information about other nodes' strategies.
We show that if they only learn their friends' strategies (with some noise),
    the system still converges close to the navigable equilibrium in a small
    number of steps.
Further, even when the node has no information about other players' strategies
    and can only use its obtained payoff as feedback to search for the best
    strategy, the system still moves close to the navigable equilibrium
    within a few hundred steps (in the $100\times 100$ grid).
Finally we simulate the DRB game on Renren and LiveJournal networks, which have non-uniform population
	distributions different from Kleinberg's grid-based small-world model.
Our simulation results show that in both networks, the game quickly converges to an equilibrium where connection preferences of users are close to
	the empirical ones. 

In summary, our contributions are the following:
    (a) we propose the small-world formation game and design a balanced
     distance-reciprocity payoff function to explain
     the navigability of real social networks;
     (b) we conduct comprehensive theoretical and empirical analysis to
     demonstrate that navigable small world is the unique robust equilibrium
     that would naturally emerge from the game under both random perturbation and strategic
     collusions; and
     (c) our game reveals a new insight between distance, reciprocity and
         navigability in social networks, which may help future research in
             uncovering deeper phenomena in navigable social
             networks.
    To our best knowledge, this is the first game theoretic
    study on the emergence of navigable small-world networks,
    and the first study that linking relationship reciprocity with network navigability.
\para{Additional related work.}
We provide additional details of prior works on explaining the emergence of navigable small-world
    networks, and other related studies not covered in the introduction.

Some studies try to explain navigability by assuming that nodes form links to optimize for
a particular property. Mathias et al. \cite{Nisha01} assume that
users try to make trade-off between wiring and connectivity. Hu et
al.~\cite{hu} assume that people try to maximize the entropy under a
constraint on the total distances of their long-range contacts.
These works rely on simulations to study the network
dynamics. Moreover, the navigability of a network is sensitive to
the weight of wiring cost or the distance constraint, and it is
    unlikely that navigable networks as defined by Kleinberg~\cite{Kleinberg00}
    would naturally emerge.

Another type of works propose node/link dynamics that converge to
    navigable small-world networks.
Clauset and Moore~\cite{Aaron03} propose a rewiring dynamic modeling a Web surfer such that
    if the surfer does not find what she wants in a few steps of greedy search, she
    would rewire her long-range contact to the current end node of the greedy search.
They use simulations to demonstrate that a network close to Kleinberg's navigable small world
    will emerge after long enough rewiring rounds.
Sandberg and Clarke~\cite{Sandberg06} propose another rewiring dynamic where with an independent
    probability of $p$ each node on a
    greedy search path would rewire their
    long-range contacts to the search target, and provide a partial analysis and simulations showing
    that the dynamic converges to a network close to the navigable small world.
Chaintreau et al.~\cite{haintreau08} use a move-and-forget mobility model, in which
    a token starting from each node conducts a random walk (move) and may also go back to the
    starting point (forget), and use the distribution of the token on the grid
    as the distribution of the long-range contacts of the starting node.
They provide theoretical analysis showing that there exists a \emph{critical} forgetting value for which the move-and-forget model
provides navigability.
However, the
underlying mechanism driving the critical value to be chosen in practice
remains unclear.

The approach taken by these studies can be viewed as orthogonal and
complementary to
    our approach: they aim at using natural dynamics (rewiring or mobility dynamics) to
    explain navigable small world, while we focus on directly exploring the entire parameter space of
    connection preferences of nodes and use game theoretic approach to show, both theoretically
    and empirically, that the nodes would naturally choose their connection preferences to form
    the navigable small world. The connection preference
can be considered a higher
level decision-making variable for individuals that pushes them to make
long-range connections over time. In particular, selecting connection preference captures the people's process of
cognitively creating behavioral plans (i.e., intensions) on how to distribute the finite time and effort
among nodes of different distance. Once the preference is selected, the players would
engage in activities such as rewiring or mobility dynamics to create long-range contacts with the corresponding connection intension.
Moreover, all the prior studies only show that they converge approximately to the navigable small
    world, while in our game the navigable small world is precisely the only robust
    equilibrium.
Finally, none of these works introduce reciprocity in their model and we are the first to link
    reciprocity with navigability of the small world.

Some studies use hyperbolic metric spaces or graphs to try to explain
    navigability in small-world networks
    (e.g. \cite{BKC09,PKBV10,KPKVB10,KPVB09,ChenFHM13,Gulyas_game}).
However, they do not explain why connection preferences in real networks
    are around the critical value and how navigable networks naturally emerge.
In particular, Chen et al.~\cite{ChenFHM13} show that the navigable small world
    in Kleinberg's model does not have good hyperbolicity.
Most recently,
    Guly\'{a}s et al.~\cite{Gulyas_game} propose a game where each player tries to minimize the
    number of links in order to be able to greedily route to all other nodes. The equilibrium of the game is
    a scale-free network whose degree distribution follows a power law. However, this game is not intended and does not explain the emergence of
    navigable small-world network validated by Milgram and others' experiments, where greedy routing can be done efficiently in $O(\log^2 n)$ time in
expectation, and relationship reciprocity is not included in any aspect of
the game.

\section{Small-world Formation Games} \label{sec:game}
In this section, we first present the game formulation based on Kleinberg's small-world model, and
we then study the payoff function which is key to understanding the underlying mechanisms that give rise to navigable small world networks.

\subsection{Game Formulation based on Kleinberg's Small-World Model}

\para{Small-world model.}Let $V = \{(i, j): i,j\in [n] =\{1, 2, \ldots, n\}\}$ be
    the set of $n^2$ nodes forming an $n\times n$ grid.
For convenience, we consider the grid with wrap-around edges connecting the
    nodes on the two opposite sides, making it a torus.
For any two nodes $u=(i_u,j_u)$ and $v =(i_v,j_v)$ on this wrap-around grid,
    the {\em grid distance} or {\em Manhattan distance} between $u$ and $v$ is defined
    as $d_M(u, v) = min\{|i_v-i_u|, n-|i_v-i_u|\}+min\{|j_v-j_u|, n-|j_v-j_u|\}$.

The Kleinberg's small-world model has two universal constants $p, q\ge 1$, such that
    (a)  each node has undirected edges connecting to all other nodes
    within lattice distance $p$, called its {\em local contacts}, and
    (b)  each node has $q$ random directed edges connecting to possibly faraway
    nodes in the grid called its {\em long-range contacts},
    drawn from the following distribution.
Each node $u$ has a {\em connection preference} parameter
    $r_u\ge 0$, such that the $i$-{th} long-range
    edge from $u$ has endpoint $v$ with probability proportional to
    $1/d_M(u,v)^{r_u}$, that is, with probability
$\chgins{p_u(v,r_u)} =d_M(u, v)^{-r_u}/c(r_u)$, where $c(r_u)=\sum_{\forall v \neq
u}d_M(u, v)^{-r_u}$ is the normalization constant.
Let $\bf r$ be the vector of $r_u$ values on all nodes.
We use ${\bf r} \equiv s$ to
    denote $r_u=s, \forall u\in V$.

    The above model can be easily extended to $k$ dimensional grid
    (with wraparound)
    for any $k = 1,2,3,\ldots$, where each long range contact is still established with
    probability proportional to $1/d_M(u,v)^{r_u}$. We use $K(n, k, p, q, {\bf r})$ to refer to the class of Kleinberg random graphs
    with parameters $n$, $k$, $p$, $q$, and $\bf r$.

\para{Small-world formation game.}A game is described by a system of players,
    strategies and payoffs.
Connection preference $r_u$ in Kleinberg's model
    reflects $u$'s intention in establishing long-range contacts:
When $r_u =0$, $u$ chooses its long-range contacts uniformly
    among all nodes in the grid;
    as $r_u$ increases, the long-range
    contacts of $u$ become increasingly clustered in its vicinity on
    the grid.
Our insight is to treat connection preference as node's strategy
    in a game setting and study the game behavior.

    More specifically, we model this via a non-cooperative game among nodes in the
    network.
First, we assume that each $r_u$ is taken from a discrete set
    $\Sigma=\{0, \gamma, 2\gamma, 3\gamma, \ldots, \}$, where
    $\gamma$ represents the granularity of connection preference and is
    in the form of $1/g$ for some positive integer $g \ge 2$.
Using discrete strategy set avoids nuances in continuous strategy space and
    is also reasonable in practice since people are unlikely to
    make infinitesimal changes.

Next, we model the small-world network formation as a
game $\Gamma = (\Sigma, \pi_u)_{u\in V}$,
    where $V$ is the set of nodes (players) in the grid, connection
    preference $r_u \in \Sigma$ is the strategy of a player $u$,
    and $\pi_u: {\cal S}\rightarrow
\mathbb{R}$ is the payoff function of $u$, with ${\cal S}=\Sigma\times
\Sigma\times\ldots\times \Sigma$.
An element $\mathbf{r}= (r_1, r_2, \ldots, r_n)\in {\cal S} $
    is called a {\em strategy profile}.

Let ${\cal C} = 2^V\setminus \emptyset$ denote the set of all coalitions.
For each coalition $C\in {\cal C}$, let $-C = V\setminus C$, and
    if $C=\{u\}$, we denote $-C$ by $-u$. We
also denote by ${\cal S}_C$ the set of strategies of players in
coalition $C$, and ${\bf r}_C$ the partial strategy profile of $\bf r$
    for nodes in $C$.

\para{Objective.}Greedy routing on the small-world network from a source node $u$ to a target
    node $v$ is a decentralized algorithm starting at node $u$, and at each
    step if routing reaches a node $w$, then $w$ selects one node from its
    local and long-range contacts that is closest to $v$ in grid distance as
    the next step in the routing path, until it reaches $v$.
In \cite{Kleinberg00}, Kleinberg shows that given a  two-dimensional grid, when ${\bf r}\equiv 2$,
    the expected number of greedy routing steps (called {\em delivery time})
    is $O(\log^2 n)$, but when ${\bf r}\equiv s \ne 2$,
    it is $\Omega(n^c)$ for some constant $c$ related to $s$.
More generally, for any $k$ dimensional grid, it is shown that ${\bf r}\equiv k$ is the
    critical value allowing efficient greedy routing.
Hence, we call Kleinberg's small world with ${\bf r}\equiv k$
    the {\em navigable small world}.

Interestingly, empirical evidences have demonstrated that the real-world network is navigable with the connection preference close to the critical value \cite{Liben-Nowell,Leskovec,Adamic,Levy,Renaud,Illenberger}. We aim to explain
this striking coincidence from the
perspective of individual incentives. In particular, our objective is to study intuitively appealing payoff functions $\pi_u$ and
    find one that individual efforts to get this payoff
lead fairly quickly to the emergence of navigable
    small-world network.

\subsection{Routing-based Payoff}
As navigable small world achieves best greedy routing efficiency,
    it is natural to consider the
    payoff function as the expected delivery time to the
    target in greedy routing.
Given the strategy profile ${\bf r} \in {\cal S}$, let
$t_{uv}(r_u,\mathbf{r}_{-u})$ be the expected delivery time from source $u$
to target $v$ via greedy routing. The payoff function is given by:
\begin{equation}
\label{eqn:rout}
\pi_u(r_u, \mathbf{r_{-u}})= - \sum_{\forall v \neq
u}{t_{uv}(r_u, \mathbf{r}_{-u})}.
\end{equation}
We take a negation on the sum of expected delivery time because nodes prefer
    shorter delivery time.

\begin{figure}[t]
\begin{minipage}[t]{0.49\columnwidth}
    \centerline{\includegraphics[width=0.8\textwidth, height =0.5\textwidth]{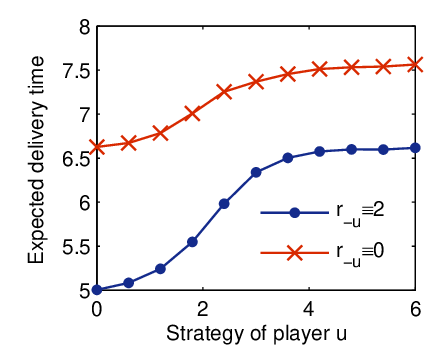}}
    \caption{The expected delivery time for a player $u$ with different strategies.}
    \label{fig:routing}
\end{minipage}
\hfill
\begin{minipage}[t]{0.49\columnwidth}
    \centerline{\includegraphics[width=0.75\textwidth, height =0.5\textwidth]{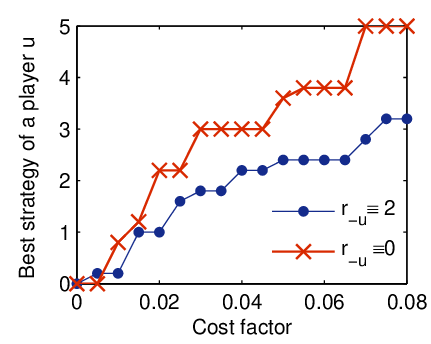}}
    \caption{The best response of a player $u$ given different cost factors.}
    \label{fig:routingbudget}
\end{minipage}
\end{figure}

Although the above payoff function is intuitive and simple, it has
    some serious issues. Prior work~\cite{Andras}
   has already proved that, with the length of greedy
paths as the payoff, player u's best response is to link uniformly
(i.e., $r_u=0$) for the one-dimensional case. For higher dimensions,
Figure~\ref{fig:routing} shows the expected delivery time for a
    single node $u$ at a $100\times 100$ grids, where each node
    generates $q=10$ links.
We see that when other nodes fixed their
    strategy (e.g., $\mathbf{r}_{-u}\equiv 2$), the best strategy of a single node $u$ is $0$.
More tests on different initial conditions reach the same result that
    the system will converge to the random small-world networks.
The intuitive reason is
that to reach other nodes quickly, it is better for a node to evenly spread its long-range
contacts from the individual prospective (or equivalently, seeking the long-range contacts of the largest distance on average given $r_u \geq 0$).
This is inconsistent with empirical evidence that
    real-world networks are navigable ones, where
links are much more likely to connect ¡°neighbor nodes¡± than
distant nodes.

In practice, creating and maintaining long-range links have higher costs, so
    one may adapt the above payoff function by adding the grid distances of
    long-range contacts as a cost term in the payoff function:
\begin{equation}
\label{eqn:rout} \pi_u(r_u, \mathbf{r_{-u}})= - \sum_{\forall v \neq
u}{t_{uv}(r_u, \mathbf{r}_{-u})}-\lambda \sum_{\forall v \neq
u}p_u(v,r_u)d_M(u,v),
\end{equation}
where $\lambda$ is a factor controlling the long range cost and $p_u(v,r_u) =d_M(u, v)^{-r_u}/c(r_u)$
is the probability that $u$ takes $v$ as a long-range contact under the strategy of $r_u$. A
larger $\lambda$ means users are more concerned with distance costs.
Figure~\ref{fig:routingbudget} shows that the best strategy of a
user $u$ is significantly influenced by the cost factor. Similar
result is also shown in~\cite{Andras}. Thus, it is unclear if the
navigable small-world network can naturally emerge from this type of
game.

In the above payoff functions, we use the expected delivery time to measure the routing efficiency to an arbitrary node.
It is also possible to give more complex payoff functions by considering the distribution
functions of delivery time, such as the percentage of nodes that can be delivered within a given number of steps.
However, given that individuals can have different strategies, it is very difficult to obtain the explicit form of delivery time $t_{uv}(r_u, \mathbf{r}_{-u})$ in terms of users strategies $\mathbf{r}$.
Due to this disadvantage of the delivery time-based games,
there is no theoretical guarantee that the network formation would converge to the
desired navigable small world.

\subsection{Distance-Reciprocity Balanced Payoff}\label{sec:DRB-payoff}
The previous section demonstrates that seeking short routing distance
    alone cannot explain the emergence of
    navigable small world, and thus people in the social network must
    have some other objective to achieve.
Reciprocity is regarded as a basic mechanism
    that creates stable social relationships in the real world~\cite{Alvin}. Several empirical studies \cite{Java,Liben-Nowell,Alan} also
show that high reciprocity is also a typical feature present in real
small-world networks (such as Flickr, YouTube, LiveJournal, Orkut
and Twitter).

Therefore, we consider the payoff of a user $u$ as the following balanced
    objective between distance and reciprocity:
\begin{align}
&\pi_u(r_u, \mathbf{r_{-u}})=  
\left(\sum_{\forall v \neq u}{p_u(v, r_u)d_M(u,v)}\right)^{\alpha_u} \times
\left(\sum_{\forall v \neq u}{p_u(v, r_u)p_v(u,r_v)}\right),
\label{eqn:balance}
\end{align}
where $\sum_{\forall v \neq u}{p_u(v,r_u)d_M(u,v)}$ is the mean
    grid distance of $u$'s long-range contacts, $\sum_{\forall v
\neq u}{p_u(v,r_u)p_v(u,r_v)}$ is the mean probability for $u$ to form
    bi-directional links with its long-range contacts, i.e., reciprocity,
    and $\alpha_u$ $(\alpha_u>0)$ is a constant exponent with respect to node $u$,
    capturing how that user weighs the relative
importance of distance and reciprocity. Note here the tradeoff exponent $\alpha_u$  could be heterogeneous among players, modeling users having different weights on the balance between
    the distance and reciprocity tradeoff. So our utility function is very flexible and actually represents a large class of tradeoff functions.
We refer the small-world formation game with payoff function in
    Eq.\eqref{eqn:balance} the {\em Distance-Reciprocity Balanced (DRB)} game.

The payoff function in Eq.\eqref{eqn:balance} reflects two natural objectives
    users in a social network want to achieve: first, they want to connect
    to remote nodes, which may give them diverse information as in the
    famous "the strength of weak ties argument" by Granovetter
    \cite{Granovetter1973strength}; second, they want to establish
    reciprocal relationship which are more stable in the long term.
However, these two objectives can be in conflict for a node $u$
    when others prefer linking in their vicinity (i.e.,
    other nodes $v$ choosing positive exponent $r_v$).
In this case, faraway long-range contacts are less likely to create
    reciprocal links.
Therefore, node $u$ should obtain the maximum payoff when it
    achieves a balance between the two objectives.
We use the simple product of distance and reciprocity objectives
    to model this tradeoff, and allow
different nodes to have different emphasis on distance-reciprocity tradeoffs with their own exponents.
One may also consider the addition of the distance term and the reciprocity term to model the tradeoff, but
	since the two quantities have different unit of scale --- distance scales from $1$ to $O(kn)$ while reciprocity is
	a probability between $0$ and $1$, we believe the multiplicative formulation makes more sense.

We remark that the reciprocity term $\sum_{\forall v \neq u}{p_u(v,r_u)p_v(u,r_v)}$
	does not consider reciprocity formed by fixed local contacts.
Effectively, we disregard local contacts and treat $p=0$ in the small
	world setting $K(n, k, p, q, {\bf r})$.
This treatment makes our analysis more streamlined and only focused on long-range
	contacts, and it also makes intuitive sense:
	the local contacts are passively given based on geographic location, while
	long-range contacts are actively established by nodes based on their connection
	preference, and thus reciprocity based on long-range contacts could make
	more sense.
For example, your neighbors in the same apartment building are your local contacts
	by physical location, but it does not mean that they are your friends, and you
	still need to intentionally establish friendship (based on your preference)
	among your neighbors, and thus reciprocity only by physical location does not
	mean much but reciprocity based on actively established relationship does
	mean a lot for an individual.

Existing network formation games typically use pure-link-based strategy and lead to mostly trivial equilibria such as cliques or stars. Different
from prior games, we use the link probability functions as the strategies, which can be viewed as a mix strategy on pure links, but with restricted distributions.
Here, we focus on the power-law distributions assumed in Kleinberg's small-world models, which is also supported from findings in several real
complex networks, such as human travel network~\cite{Marta,Kai}, communication network~\cite{Krings}, trade network~\cite{Bhattacharya} and other
social networks~\cite{Liben-Nowell,Leskovec}.

In practice, the strategy captures certain behavioral preference of players related to connection.
One concrete example is mobility preference in human travel network, where the link distribution can be understood as the trip distance distribution by taking the grid location of a node as its home.
In this network, the strategy $r_u$ captures the mobility preference of individual $u$, large $r_u$ results in large possibility of short
distance travel. Our game means that each user adjusts its mobility preference to the heterogeneous preferences of others for a better payoff, such as obtaining non-redundant information via long-distance visits and social support enforced by mutual visits.

\section{Properties of the DRB Game} \label{sec:theory}
In this section, we conduct theoretical analysis to
    discover the properties of the DRB game. We begin by considering the problem of the existence of equilibria in the game, and if the answer is yes, whether there exist multiple equilibria. In Section~\ref{sec:existence}, we prove that DRB game has {\em only two} Nash equilibria
    ${\bf r}\equiv k$ and ${\bf r} \equiv 0$, corresponding to the navigable
    and random small-world networks, respectively.
    Given multiple
	Nash equilibria, we further investigate if
   the navigable small world possesses further properties making it the likely
   choice in practice. This is the task of the next two sections.

One way to solve the problem of multiple equilibria is to consider a more appealing equilibrium concept--strong Nash equilibrium (SNE). While in a NE no player can
improve its payoff by unilateral deviation, in a SNE there is no coalition
of players that can improve their payoffs by collective deviation. In Section~\ref{sec:collusion}, we show that the navigable small-world equilibrium is a SNE in the game, which is much more stable than the random small-world equilibrium.

Another
way to approach the problem is to study the convergence to equilibrium under
the best response dynamics. This dynamics could help to
select among multiple equilibria of the game. In Section~\ref{sec:perturb}, we show that the navigable small-world equilibrium is reachable via best response dynamics from
any state not in the other equilibrium. We also prove that the navigable small-world NE can also tolerate large perturbations of players under best response dynamics, whereas the random small-world NE is extremely unstable under perturbation.

We finally give a description how the
navigable small-world network is formed by summarizing our results in Section~\ref{sec:implication}.

\subsection{Equilibrium Existence}\label{sec:existence}
{\em Nash equilibrium (NE)} for the strategic game $\Gamma = (\Sigma, \pi_u)_{u\in V}$ is a strategy
profile $\mathbf{r}^* \in {\cal S}$ such that each player's strategy $r_u^*$ ($\forall u\in V$) is a best response to the
other players' strategies $\mathbf{s}^*_{-u}$, where the best response is defined as follow:
\begin{definition} [Best response]Player $u$'s strategy
$r^*_u \in \Sigma $ is a best response to the strategy profile
$\mathbf{r_{-u}}\in {\cal S}_{-u}$ if
\begin{displaymath}
\pi_u(r^*_u,\mathbf{r_{-u}}) \geq \pi_u(r_u,\mathbf{r_{-u}}),
\forall r_u \in \Sigma \setminus\{r^*_u\},
\end{displaymath}
Moreover, if ``\,$\ge$'' above is actually ``\,$>$'' for all $r_u\ne r_u^*$,
    then $s_u^* $ is the unique best response to $\mathbf{s_{-u}}$. We denote this unique best response as $B_u(\mathbf{s}_{-u})$. Strategy profile $\mathbf{r}^*$ is a strict Nash equilibrium
    if for every player $u\in V$, $r_u^*$ is the unique best response to $\mathbf{r}^*_{-u}$.
\end{definition}

We first show that the navigable small-world network is a Nash Equlibrium
of the DRB game. To do so, we
focus on a local region centering around a node $w$ preferring local connection,
and we have the following important lemma.
\begin{restatable}{lem}{neighbors}
\label{lem:neighbors}
In the $k$-dimensional DRB game, for any constant $\delta$,
	there exists $n_0\in \mathbb{N}$ (may depend on $\delta$),
	for any $n\ge n_0$,
	for any non-zero strategy profile ${\bf r} \not \equiv 0$,
	if a node $w$ satisfies
	$r_w \geq k$ or $r_w = \max_{v\in V} r_v$,
	then for any node $u$ within $\delta$ grid distance of $w$ (i.e. $d_M(u,w) \leq \delta$),
	$u$ has the unique best response of $r_u=k$.
\end{restatable}
\begin{proof}[(Sketch)]
The intuition is as follows. When a node $w$ satisfying
$r_w \geq k$ or $ r_w = \max_{v\in V} r_v$, it prefers its
	long-range contacts to be in its vicinity.
For a nearby node $u$ with $d_M(u,v)\leq \delta$, the case of $r_u=k$ provides
	the best balance between good grid distance
    to long-range contacts and high reciprocity (even just counting the
    reciprocity received from $w$). In other cases,
    the node $u$ obtains either too low reciprocity or too short average grid distance to long-range contacts.
    In the case of $r_u < k$,
    the node $u$ could increase the average grid distance to long-range contacts by an factor of $O(\ln n)$, but the reciprocity can be reduced by a factor of  $\Omega(n^{\gamma})$, as compared with those provided by $r_u=k$. Thus, the ratio of payoff for $r_u< k$ to payoff for $r_u=k$ is at most $O(\ln^{\alpha_u} n/ n^{\gamma})$, which is smaller than one given sufficient large $n$. Similarly, in the case of $r_u > k$,
    the node $u$ could increase the reciprocity by an factor of $O(\ln n)$, but the average grid distance to long-range contacts can be reduced by a factor of  $\Omega(n^{\gamma})$, as compared with those provided by $r_u=k$. Thus, the ratio of payoff for $r_u> k$ to payoff for $r_u=k$ is also at most $O(\ln n/ n^{\alpha_u\gamma})$, which is also smaller than one given sufficient large $n$.
     The detailed proof is included in Appendix~B.
\end{proof}
The above lemma shows that given a non-zero profile, we can find a
local region where the best response of every node is $k$.
This lemma is instrumental to several analytical results, including the following theorem.

\begin{restatable}{thm}{response} For the DRB game in a $k$-dimensional grid,
the following is true
    for sufficiently large $n$:
\footnote{Technically, a statement being true
    for sufficiently large $n$ means that there exists a constant
    $n_0\in \mathbb{N}$ that may only depend on model constants such
    as $k$, $\gamma$ and $\alpha_u$, such that for all $n\ge n_0$ the statement is true
    in the grid with parameter~$n$.}
For every node $u \in V$, every strategy profile $\bf r$,
    and every $s \in \Sigma$, if $\mathbf{r}_{-u} \equiv s$, then
    $u$ has a unique best response to ${\bf r}_{-u}\equiv s$:
\begin{displaymath}
    B_u({\mathbf{r}_{-u}} \equiv s)=
   \begin{cases}
   k & \mathrm{if}\;  s>0,\\
   0 & \mathrm{if}\;  s=0.
   \end{cases}
\end{displaymath}
\label{thm:balancedbestk}
\end{restatable}\vspace{-2mm}
\begin{proof}[(Sketch)]
For the case of $s>0$, given the strategy profile of $\mathbf{r}_{-u} \equiv s$,
for every node $u$, each of its nearest neighbor $w$ (i.e., $d_M(u,w)=1$) satisfies $r_w = \max_{v\in V} r_v$.
Thus by Lemma~\ref{lem:neighbors}, node $u$'s unique best response to ${\bf r}_{-u}\equiv s$ is $r_u=k$.

When $s=0$, all others nodes link uniformly. In this case, the
reciprocity for node $u$ becomes a constant independent of its
strategy $r_u$. Thus, $r_u$ should be selected to maximize average
distance of $u$'s long-range contacts, which leads to $r_u=0$. 
The detailed proof of this case is
included in Appendix~C.
\end{proof}

Theorem~\ref{thm:balancedbestk} shows that when all other nodes use
the same nonzero
    strategy $s$, it is strictly better for $u$ to use strategy $k$; when all other nodes
    uniformly use the $0$ strategy, it is strictly better for $u$ to also use $0$ strategy.
When setting $s=k$ and $s=0$, we have:
\begin{restatable}{cor}{corNE} \label{cor:NE}
For the DRB game in the $k$-dimensional grid, the navigable
    small-world network (${\bf r} \equiv k$) and the random small-world network (${\bf r}\equiv 0$)
    are the two strict Nash equilibria for sufficiently large $n$, and there
    are no other uniform Nash equilibria. %
\end{restatable}


We next examine if there exists any non-uniform equilibrium.
\begin{restatable}{thm}{uniqueness}
In the $k$-dimensional DRB game, there
is no non-uniform Nash equilibrium for sufficiently large $n$.
\label{thm:uniqueness}
\end{restatable}
\begin{proof}
Given any non-uniform strategy profile ${\bf r}$, let $V_{\geq k}=\{v| r_v \geq k\}$.
If $V_{\geq k} \neq \emptyset$,
we can find a pair of grid neighbors $(u,w)$ with $r_u \neq k$ and
$r_w\ge k$.
If $V_{\geq k} = \emptyset$, we can find a pair of grid neighbors $(u,w)$
	with $r_u \neq k$ and
$r_w = \max_{v\in V} r_v$. In either case, we know the node $u$
could obtain better payoff
by unilaterally deviating to the strategy $r_u=k$ by Lemma~\ref{lem:neighbors}.
Therefore, non-uniform strategy profile ${\bf r}$ is not a Nash equilibrium.
\end{proof}
Combining the above theorem with Corollary~\ref{cor:NE}, we see that DRB game has {\em only two} Nash equilibria
    ${\bf r}\equiv k$ and ${\bf r} \equiv 0$, corresponding to the navigable
    and random small-world networks, respectively.

\subsection{Equilibrium Stability under Collusion}\label{sec:collusion}
While in an NE no player can improve its payoff by unilateral
deviation, some of the players may benefit (sometimes substantially)
from forming alliances/coalitions with other players. So we study a more general \emph{$t$-Strong Nash equilibrium ($t$-SNE)} to study the resilience to coalitions.

\begin{definition}[$t$-Strong Nash equilibrium] For a number $t \in \{1,2,\ldots, |V|\}$,
a strategy profile $\mathbf{r}^* \in
{\cal S}$ is a $t$-strong Nash equilibrium if for all $ C\in {\cal C}$ with
    $|C| \le t$,
there does not exist any $r_C \in {\cal S}_C$ such that
\begin{displaymath}
\forall u\in C, \pi_u(\mathbf{r}_C, \mathbf{r}^*_{-C})\geq \pi_u(\mathbf{r}^*),
\exists u\in C, \pi_u(\mathbf{r}_C, \mathbf{r}^*_{-C}) > \pi_u(\mathbf{r}^*). \end{displaymath}
When $t=|V|$, we simply call $\mathbf{r}^*$ the {\em strong
Nash equilibrium (SNE)}. Note that $1$-SNE falls back to NE.
\end{definition}


    We first show the important result that the navigable small-world
    network is able to tolerate collusion of any group
    of players, i.e., ${\bf r}\equiv k$ is a $|V|$-SNE or simply SNE.

\begin{restatable}{thm}{strongNE}
For the DRB game in the $k$-dimensional grid,
    the navigable small-world network (${\bf r}\equiv k$) is a strong Nash equilibrium for sufficiently large $n$.
 \label{thm:strongNE}
\end{restatable}
\begin{proof}[(Sketch)]
We prove a slightly stronger result --- any node $u$ in any strategy
    profile $\bf r$ with $r_u\ne k$ has
    strictly worse payoff than its payoff in the navigable small world.
Intuitively, when $u$ deviates to $0 \leq r_u<k$, its loss on
reciprocity would outweigh its gain on link distance;
    when $u$ deviates to $r_u>k$, its loss on link distance is too much
    to compensate any possible gain on reciprocity.
The detailed proof is in Appendix~D.
\end{proof}

The above theorem shows that the navigable small-world equilibrium is not only
immune to unilateral deviations, but also to deviations by
coalitions of any size, and in particular it is \emph{Pareto-optimal}, such that
    no player can improve her payoff without decreasing the
    payoff of someone else.

After showing that the navigable small-world is robust to collusions of any size, we now
show that random small world equilibrium is not stable even under the collusion of
a pair of nodes.
\vspace{-2mm}
\begin{restatable}{thm}{coalition}
For the DRB game in a $k$-dimensional grid,
    the random small-world NE $\mathbf{r}\equiv 0$ is not
   a $2$-strong Nash equilibrium for sufficiently large $n$.
    \label{thm:coalition}
\end{restatable}
\begin{proof}[(Sketch)]
If a pair of grid neighbors
collude to deviate their strategies to $k$, they could gain much benefit in terms of reciprocity,
as compared with the loss of relationship distance. As a result, they would both get better payoff than their payoff in $\mathbf{r}\equiv 0$.
The detailed proof is in Appendix~E.
\end{proof}

\subsection{Convergence under Best Response Dynamics}\label{sec:perturb}
For our game, we finally study its best response dynamics to investigate its properties
	of convergence to Nash equilibria.
Best response dynamics are typically specified in terms of {\em asynchronous steps}:
	in each asynchronous step, {\em one} player moves from its current strategy to its
	best response to the current strategy profile, and thus the entire strategy profile
	moves one step accordingly.
To facilitate the study of convergence speed, we also look into
	{\em synchronous steps} for the best response dynamics:
	in each synchronous step, {\em every} player moves from its current strategy
	to its best response to the current strategy profile, and collectively we count
	this as one synchronous step.

With the concept of best-response dynamics, we first show that for any non-zero profile, we can find a node
    that triggers a cascade of adopting strategy $k$ from neighbors to neighbors of neighbors,
    and so on, ultimately leading to the navigable small world equilibrium.

\begin{restatable}{thm}{reachable}  \label{thm:reachable}
In the $k$-dimensional DRB game, for sufficiently large $n$,
the navigable small-world equilibrium ${\bf r}\equiv k$
is reachable via best response dynamics
	from any non-zero strategy profile ${\bf r} \not \equiv 0$.
Moreover, if all nodes move synchronously in the best response dynamics, then
	it takes at most $k\lfloor n/2\rfloor$ synchronous
	steps for any non-zero strategy profile
	to converge to the navigable small-world equilibrium ${\bf r}\equiv k$.
\end{restatable}

\begin{proof}
Let $V_w(j) = \{v|d_M(v,w) \leq j\}$. Given a non-zero profile ${\bf r}$,
we can find a node $w$ satisfying $r_w \geq k$ or $r_w = \max_{v\in V} r_v$.
Given a constant $\delta$ $(\delta \ge 2)$, Lemma~\ref{lem:neighbors} implies
	that for sufficiently large $n$,
	for every $u \in V_w(\delta)$, in one asynchronous step
	$u$ will set $r_u=k$.
Then consider $u$'s neighbors $V_u(\delta)$, in one asynchronous step each of them
	will also set their strategy to $k$.
Following this cascade it is clear that there exists a step sequence such that
	the non-zero profile ${\bf r}$ will reach the navigable small world
	${\bf r}\equiv k$.

We now consider that all nodes move synchronously.
Again we first find a node $w$ satisfying $r_w \geq k$ or $r_w = \max_{v\in V} r_v$.
By Lemma~\ref{lem:neighbors} all nodes in $V_w(\delta/2)$ ($\delta/2\ge 1$)
	move to strategy $k$ in the first synchronous step.
Consider the second synchronous step.
Even though we are not sure if node $w$ adopts strategy $k$ in the first synchronous
	step, we know that $w$ adopts $k$ in the second synchronous step since
	$w$ has neighbors adopting $k$ after the first synchronous step.
Moreover, for all nodes in $V_w(\delta/2)$, their mutual grid distance is at most
	$\delta$, and thus Lemma~\ref{lem:neighbors} applies to these nodes
	in the second synchronous step and they all stay at strategy $k$.
Finally for their grid neighbors within grid distance $\delta/2$,
	essentially nodes in $V_w(\delta)\setminus V_w(\delta/2)$, they will also
	adopt strategy $k$ in the second synchronous step.
Repeating the above procedure, all nodes that have adopted $k$ will keep $k$ while
	their grid neighbors will also adopt $k$.
Since the longest grid distance among nodes in the
	$k$-dimension grid is $k\lfloor n/2\rfloor$,
	after at most $k\lfloor n/2\rfloor$ synchronous steps, all nodes adopt $k$.
\end{proof}

The proof of the above theorem provides valuable insights into the scalability of the game.
Notice that a $k$-dimension grid contains a total of $|V|=n^k$ players, so the above theorem
states that, for any non-zero strategy profile, the convergence time to navigable NE is at most $O(|V|^{\frac{1}{k}})$ synchronous
	steps if players move synchronously in the best response dynamics. Also, any player $u$ involved in the cascade of adopting $r_u=k$
can make this best decision locally according to the strategies of the players in his
neighborhood. Thus our game is scalable with the number of players.

Next, we would like to see if the navigable equilibrium can also tolerate
    perturbations of players under best response dynamics, where the perturbations could be
    arbitrary and there is no guarantee that perturbed players are better off.
From Theorem~\ref{thm:reachable},
	we know that as long as not all nodes deviate to zero,
	there exists a best response dynamic sequence for the system to go back to
	the navigable small world, and if all nodes move synchronously, the system
	reaches the navigable small world in at most $k\lfloor n/2\rfloor$
	synchronous steps.
We now give a further result on the stability of navigable small-world
	in tolerating perturbations of random players:
	we show that even if each individual independently perturbs to an arbitrary strategy
	with a fairly large probability, the system
	moves back to the navigable small world in just one synchronous step, and
	even if players move asynchronously, it is guaranteed that the system
	moves back to the navigable small world after each node takes at least one
	asynchronous step.

\begin{restatable}{thm}{perturbNE}   \label{thm:perturbNE}
Consider the navigable small-world equilibrium ${\bf r}\equiv k$
	for the DRB game in a $k$-dimensional grid $(k>1)$.
Suppose that with probability $p_u$ each node $u\in V$ independently perturbs
	$r_u$ to an arbitrary strategy $r'_u \in \Sigma$, and with probability
	$1-p_u$ $r'_u=r_u$. Let $\alpha_{min} = \min_{u \in V}\alpha_u$,
 then for any constant $\varepsilon$ with $0<\varepsilon<\min\{1,\alpha_{min}\}\gamma/4$,
    there exists $n_0 \in \mathbb{N}$ (depending only on $k$, $\gamma$,
    and $\varepsilon$), for all $n\ge n_0$,
    if $p_u \le 1- n^{-\varepsilon}$, with probability at least $1-1/n$,
    the perturbed strategy profile ${\bf r}'$ moves back to the navigable
    small world (${\bf r}\equiv k$) in one synchronous step, or as soon as
    every node takes at least one asynchronous step in the best response dynamics.
\end{restatable}
\begin{proof}[(Sketch)]
The independently selected deviation node set satisfies
    that with high probability,
    for any node $u$, at sufficiently many distance levels from $u$
    there are enough fraction of non-deviating nodes.
We then show that $u$ obtains higher order payoff just from these
    non-deviating nodes than any possible payoff she could get from
    any possible deviation.
The detailed proof is in Appendix~F.
\end{proof}

Notice that the bound of $1- n^{-\varepsilon}$ is close to $1$ when $n$ is sufficiently large,
    meaning that the navigable equilibrium tolerates
    arbitrary deviations from a large number of random nodes.

For the random small-world network, which is shown to be the other NE,
Theorem~\ref{thm:reachable} already implies that
even one deviating player could possibly drive the system out of the
	random small-world equilibrium and lead it towards the navigable small-world equilibrium.
However, converging to navigable small world is not guaranteed in this case.
In the following, we show a stronger convergence result:
	if each individual $u$ deviates from $r_u=0$ independently with even a small probability,
	then the system could switch to the navigable small world in just one synchronous step,
	or after each node takes at least one asynchronous step, and the convergence to
	the navigable small world is guaranteed in this case.
\begin{restatable}{thm}{perturbzero}  \label{thm:perturbzero}
For the DRB game in a $k$-dimensional grid $(k>1)$ with
    the initial strategy profile ${\bf r}\equiv 0$ and a finite perturbed
    strategy set $S \subset \Sigma$ with at least one non-zero entry
    ($0< \max{S}\leq \beta$),
    for any constant $\varepsilon$ with $0<\varepsilon<\gamma/2$,
    there exists $n_0 \in \mathbb{N}$ (depending only on $k$, $\gamma$,
        and $\varepsilon$), for all $n\ge n_0$,
    if for any $u\in V$, with independent probability of $p \geq
    n^{-\frac{(k-1)\varepsilon}{k+\beta}}$,
   $r_u \in S \setminus \{0\}$ after the perturbation,
    then with probability at least $1-1/n$,
    the network converges to the navigable small world in one synchronous step, or
    as soon as every node takes at least one asynchronous step in the best response dynamics.
\end{restatable}
\begin{proof}[(Sketch)]
We consider the gain of a node $u$ when selecting $r_u=k$ separately from
    each group of nodes with the same strategy after the perturbation,
    and then apply the results in Theorem~\ref{thm:balancedbestk}.
The full proof is in Appendix~G.
\end{proof}

Note that $1/n^\frac{(k-1)\varepsilon}{k+\beta}$ is very small for
large $n$ and a finite perturbed
    strategy set $S$, which implies that the best
response of any node $u$ in the perturbed profile becomes $r_u=k$ as
long as a small number of random nodes are perturbed to a finite set
of nonzero strategies.

\subsection{Implications from Theoretical Analysis} \label{sec:implication}
Combining the above theorems together, we obtain a
better understanding of how the navigable small-world network is formed.
From any arbitrary initial state, best response dynamic drives the system
    toward some equilibrium, with the navigable small world as one of them
    (Corollary~\ref{cor:NE} and Theorem~\ref{thm:reachable}).
Even if the systems temporarily converges to a non-navigable equilibrium,
    the state will not be stable --- either a small-size collusion
    (Theorem~\ref{thm:coalition}) or a small-size random perturbation
    (Theorem~\ref{thm:perturbzero}) would make the system leave the current
    equilibrium and quickly enter the navigable equilibrium.
Once entering the navigable equilibrium, it is very hard for the system
    to move away from it --- no collusion of any size would
    drive the system away from this equilibrium (Theorem~\ref{thm:strongNE}),
    and even if a large random portion of nodes deviate arbitrarily the system
    still converge back to the navigable equilibrium as long as each node takes one
    best-response step
    (Theorem~\ref{thm:perturbNE}).
These theoretical results strongly support that the navigable
    small world is the unique stable system state, which suggests that
    the fundamental balance between reaching out to remote people and
    seeking reciprocal relationship is crucial to the emergence of
    navigable small-world networks.

\section{Quality of Equilibria}
In a Nash equilibrium, each user is maximizing its individual payoff.
However, there is also a global function of social welfare, which is the total payoff
of all nodes. A natural question then is how the social welfare
of a system is affected when its users are
selfish. Thus, in this section, we would like to examine how good the solution
	represented by an equilibrium is relative to the global optimum.

%
To study the social welfare, we
focus on the homogenous network in which all players use the same tradeoff exponent $\alpha$, since it is difficult to normalize and integrate individual utility measures if they have different emphasis on distance or reciprocity. We first examine the global optimum.
\begin{restatable}{thm}{optimal}
In the $k$-dimensional homogeneous DRB game, the optimal social welfare is $\Theta\left(\frac{n^{\alpha+k}}{\ln^{\alpha+1} n} \right)$ for sufficiently large $n$.
 \label{thm:optimal}
\end{restatable}
\begin{proof}[(Sketch)]
We prove that a node $u$ with $r_u=k$ could get high payoff if it has at
least one neighbor $v$ with $r_v > k$. In this case, the node $u$ could
get both large grid distance to long-range contacts and high reciprocity
(at least from $v$). So if the system has a constant fraction of such nodes,
the social welfare is optimized. The detailed proof is included in Appendix~H.
\end{proof}
The proof of the above theorem provides some interesting insights:
First, the optimal strategy profile is not the navigable network where all players get the same payoff, instead, it exhibits \emph{inequality} in the distribution
of payoff. The rich (e.g., those with strategy of $k$) could get high payoff
whereas the poor (e.g., those with strategy larger than $k$) only get very low payoff.
Furthermore, the optimum social welfare is achieved when the poor sacrifice their
	distance payoff and focus on their reciprocity (by selecting a strategy greater than $k$),
	so that their rich neighbors could obtain a high balanced payoff of both distance and
	reciprocity.
This situation reminds us social relationships generated by different
	social status (e.g. employee-employer relationship) or by tight bonds with
	mutual understanding and support (such as marriage relationship).

We next focus on the standard measures of the
sub-optimality introduced by self-interested behavior. In particular,
price of stability (PoS) is the ratio of the
solution quality at the best Nash equilibrium relative
to the global optimum, whereas the price of anarchy (PoA) is the ratio of the worst
Nash equilibrium to the optimum.
\begin{restatable}{thm}{worst}
In the $k$-dimensional homogeneous DRB game, for sufficiently large $n$, the PoS
is $\Theta(\ln n)$ and the PoA is $\Theta\left(\frac{n^k}{\ln^{\alpha+1} n}\right)$.
 \label{thm:worst}
\end{restatable}
\begin{proof}[(Sketch)]
From the analysis in Section~\ref{sec:existence}, we know that the system has only two
Nash equilibria ${\bf r}\equiv k$ and ${\bf r} \equiv 0$, corresponding to navigable
and random small-world networks, respectively. We show that the navigable
small-world NE is a better equilibrium since the strategy of $k$ provides
the best balance between grid distance
to long-range contacts and reciprocity. Combined with Theorem~\ref{thm:optimal},
we get the PoS and PoA of the system.
The detailed proof is included in Appendix~I.
\end{proof}
The above theorem indicates that, in the good case when the system is in the navigable network
	equilibrium, the social welfare is reasonably close to the social optimum
	(with ratio $\Theta(\ln n)$ among $n^k$ nodes), but in the bad case when the network
	is in the random network equilibrium, the social welfare is far from the social optimum.


\section{Empirical Evaluation} \label{sec:evaluation}
In this section, we empirically examine the stability of navigable
small-world NE.
We simulate the DRB game on two dimensional grids, and consider
    nodes having full information, limited information, or no information
    of other players' strategies.


In Section~\ref{sec:perturb_eval} and Section \ref{sec:part_eval}, we focus on the homogeneous game ($\alpha_u=1$, $\forall u \in V$)
 as our equilibrium analysis is robust to $\alpha_u$ under the $k$-dimensional grid of people.
 In Section \ref{sec:real_eval}, we further examine the heterogeneous game under non-uniform
 population density across real social networks.
 Before the main empirical evaluation, we first test the effect of
    the grid size $n$ on navigable equilibrium, since our theoretical results
    require a sufficiently large $n$. 

    Our theoretical analysis shows that one can find a large enough
    constant $n_0$, such that the navigable equilibrium is exactly
    ${\bf r}\equiv 2$ for all $n\geq n_0$.
Thus, we first verify empirically the relationship between the size of the
    grid and the actual connection preference value for the
    equilibrium.
Figure~\ref{fig:enquilibrium} shows how the equilibrium value
changes over $n$ in a 2D grid, with various granularity. For example, with a granularity of $\gamma=0.1$,
the equilibrium decreases from $\mathbf{r} \equiv
2.3$ for a very small $10\times 10$ grid, to $\mathbf{r} \equiv
2$ for a $1000\times 1000$ grid. This shows that we do not need a
very large grid in order to obtain
    results close to our theoretical predictions.
In our following experiments, we use a
$100\times 100$ grid with the granularity $\gamma=0.1$, which leads to an equilibrium $\mathbf{r} \equiv 2.1$ close to
theoretical prediction while reducing the simulation cost.

\subsection{Stability of NE under Perturbation}\label{sec:perturb_eval}

To demonstrate the stability of navigable NE, we simulate the DRB
game with random perturbation. At time step $0$, each player is
perturbed independently with probability $p$. If the perturbation
occurs on a player $u$, we assume that the player $u$ chooses a new
strategy uniformly at random from the interval $[0, 10]\cap \Sigma$.
Notice that for strategy $r_u>10$, the behavior of nodes is similar
to $r_u=10$ as nodes only connect to the 4 grid neighbors. 
Let ${\bf r}^0$ be the strategy profile at time $0$ after the
perturbation. At each time step $t \geq 1$, every player picks the
best strategy based on the strategies of others in the previous
step: $r_u^t=\argmax_{r_u \in \Sigma\cap [0,10]} \pi
(r_u,\mathbf{r}^{t-1}_{-u}), \forall u, \forall t
>1.$

\begin{figure}[t]
\begin{minipage}[t]{0.32\columnwidth}
    \centerline{\includegraphics[width=1\textwidth, height =0.85\textwidth]{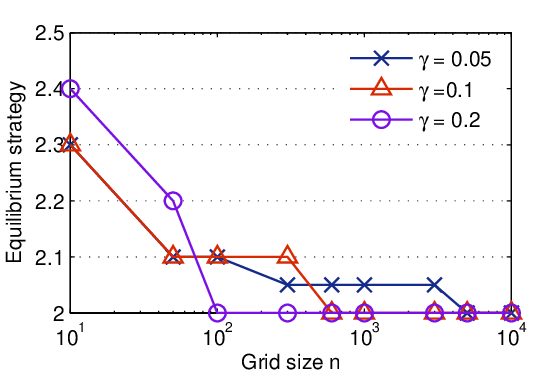}}
    \caption{Equilibrium strategy in the 2D grid of different size and granularity.}
    \label{fig:enquilibrium}
\end{minipage}
\hfill
\begin{minipage}[t]{0.32\columnwidth}
    \centerline{\includegraphics[width=1\textwidth, height =0.8\textwidth]{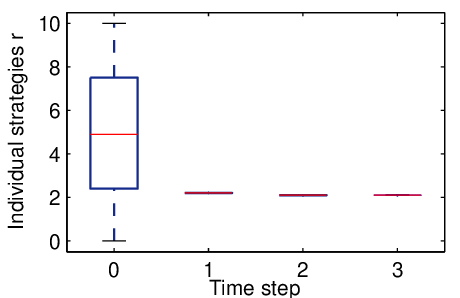}}
    \caption{The return to navigable small-world NE (perturbed
probability p=1).}
    \label{fig:SNE_perturb}
\end{minipage}
\hfill
\begin{minipage}[t]{0.32\columnwidth}
    \centerline{\includegraphics[width=1\textwidth, height =0.8\textwidth]{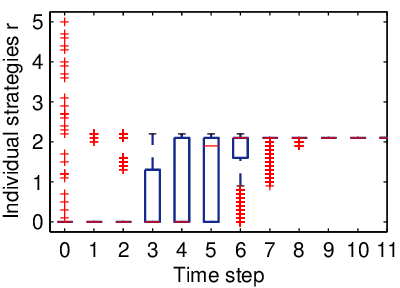}}
    \caption{From random NE to small-world NE (perturbed
probability p=0.01).}
    \label{fig:RNE_perturb}
\end{minipage}
\end{figure}

Figure~\ref{fig:SNE_perturb} shows an extreme case where every player is
    perturbed when the initial profile is ${\bf r} \equiv 2$.
The box-plot shows the distribution of players' strategies at each step.
The figure shows that in just two steps the system returns to the
    navigable small-world NE. We tested $100$ random starting profiles,
    and all of them converge to the navigable NE within two steps. This simulation
result indicates that the navigable NE is very stable for random
perturbations.


To contrast, we study the stability of the random small-world network
    in terms of tolerating perturbations.
Figure~\ref{fig:RNE_perturb} shows the result of randomly perturbing only
    $1\%$ of players at the random NE, which are shown as the
    outliers at step $0$.
Note that $1\%$ perturbation does not meet the requirement in
    Theorem~\ref{thm:perturbzero}.
However, this small fraction of players would affect the
decision of additional players in their vicinity, who can
significantly improve the reciprocity by also linking in the
vicinity (indicated by Theorem~\ref{thm:reachable}).
The figure clearly shows that in a few steps,
    more and more players would change their strategies,
    and the system finally goes to the navigable
    small-world NE.\footnote{In step 1 and 2 in Figure~\ref{fig:RNE_perturb},
    the number of outliers is larger than in step 0, even though the rendering
    make it seems they are less.}
We tested $100$ random starting profiles, and all of them
converge to the navigable NE within at most 12 steps.

These results show that the navigable small-world NE are robust to
perturbations, while random small-world NE is not stable and
    easily transits to the small-world NE under a slight perturbation.

\subsection{DRB Game with Limited Knowledge}\label{sec:part_eval}
In practice, a player does not know the strategies of all players.
So we now consider how to operate best response dynamics in practical scenarios.
We first examine a weaker scenario where a player only knows the
strategies of their friends.
With these limited knowledge, a player can guess the strategies of
all other players and pick the best response to the estimated
strategies of all players.
We next consider the
weakest scenario where each player has no knowledge about the
strategies of other players, and the only information needed is the empirical payoff observed by the player. To get this information, a
player can create a certain number of links with the current strategy,
and compute the payoff by multiplying the average link distance and
the percentage of reciprocal links. In this scenario, players cannot directly calculate the best responses.
Instead, they perform a heuristic search through choosing a response of better payoff than their current strategies,
whenever they have opportunities to adjust the strategies. So as the time goes on, the player could change the strategy
towards the best response.

\para{Scenario 1: knowing friends' strategies.}To examine the convergence of navigable small-world NE in this
scenario, we simulate the DRB game as follows.
 At time step
$0$, each player chooses an initial strategy uniformly at random
from the interval $[0, 10]\cap \Sigma $. At every step $t\ge 0$,
each player $u$ creates $q$ out-going
    long-range links based on her current strategy $r_u^t$, and learns
    the connection preferences of these $q$ long-range contacts.
Let $F_u^t$ be the set of these $q$ long-range contacts.
We further allows a random noise term $\varepsilon$ for each
    connection preference learned from the friends.
Let $\hat r_v^t$ ($v\in F_u^t$) be the learned (noisy) connection preference.
Then based on these newly learned connection preferences,
    player $u$ estimates the strategies of all other
    players.
One reasonable estimation method is to assume that players close to one another
    in grid distance have similar strategy.
More specifically, for a non-friend node $v \not\in F_u^t$, $u$ estimates
    the strategy of $v$ by the average weight of known strategies: $\hat r_v^t= \frac{\sum_{f\in F_u^t}{\hat r_{f,t-1}/d_M(v,f)}}{\sum_{f\in
    F_u^t} 1/d_M(v,f)}.$

\begin{figure*}[t]
  \centering
  \begin{minipage}[t]{0.32\columnwidth}
    \includegraphics[width=1\textwidth, height =0.8\textwidth]{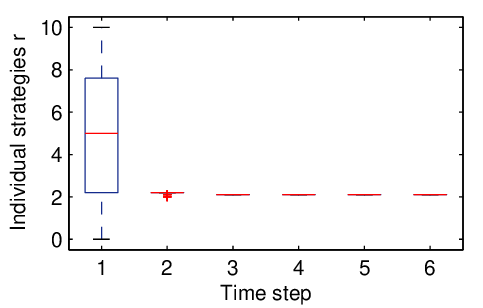}
    \caption{Player only knows the strategies of
      their friends (Noise $\varepsilon=0$).}
    \label{fig:estimate:a}
  \end{minipage}
  \begin{minipage}[t]{0.32\columnwidth}
      \includegraphics[width=1\textwidth, height =0.8\textwidth]{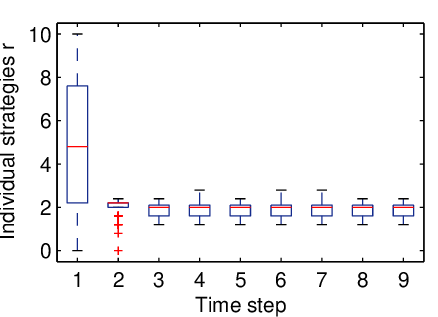}
      \caption{Player only knows the strategies of
      their friends (Noise $\varepsilon\sim N(0,0.5)$)}
      \label{fig:estimate:b}
  \end{minipage}
  \hfill
  \label{fig:estimate}
  \begin{minipage}[t]{0.32\columnwidth}
    \centerline{\includegraphics[width=1\textwidth, height =0.8\textwidth]{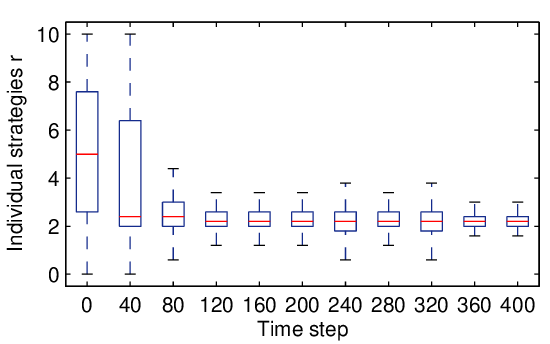}}
    \caption{Players have no knowledge of strategies of others.}
    \label{fig:NE_arrival}
  \end{minipage}
\end{figure*}

Here we do not use the connection preferences learned in the previous steps and
    effectively assume that those old links are removed.
This is both for convenience, and also reasonable since people could only maintain
    a limited number of connections and it is natural that new connections
    replace the old ones.
Moreover, the connection preferences of those old connections
    may become out-dated in practice anyway.
After the estimation procedure, player $u$ uses the strategy $\hat r_v^t$ from
    all other players (either learned or estimated) to compute
    its best response $\hat r_u^{t+1}$ for the next step.


In our experiment, we set $q=30$.
Figure~\ref{fig:estimate:a} shows that when players have accurate
    knowledge of the strategies of their friends without noise,
    the system converges in just two steps.
Even when the information on friends' strategies is noisy,
    the system can still quickly stabilize in a few steps
    to a state close to the navigable small-world NE, as shown in
    Figure~\ref{fig:estimate:b}. We tested $100$ random starting profiles
    and also other estimation methods such as randomly choosing a connection
    preference based on friends' connection preference distributions,
    and results are all similar.
This experiment further demonstrates the robustness of the small-world NE
    even under limited information on connection preferences.

\para{Scenario 2: No information about others' strategies.}To make it even harder, we do not allow the player to try many different
    strategies at each step before fixing her strategy for the step.
Instead, at each step each player only has one
    chance to slightly modify her current strategy.
If the new strategy yields better payoff, the player would adopt the new
strategy. So as the time goes on, the player could change the
strategy towards the best one.


We simulate the DRB game as follows: At time step $0$, each player
chooses an initial strategy uniformly at random from the interval
$[0, 10] \cap \Sigma $. Every player creates $q$ out-going links with her current
strategy. At each time step $t \geq 1$, each player changes the
strategy, i.e.,$r_u\gets r_u+\delta$, and creates $q$ new links with
this new strategy, where $\delta$ is a random number determined as follows.
First, for the sign of $\delta$, in the first step it is randomly assigned positive or negative
    sign with equal probability; in the remaining steps, to make the search efficient,
        we keep the sign of $\delta$ 
        if the previous change leads to a higher payoff; otherwise
        we reverse the sign of $\delta$. 
For the magnitude of $\delta$, i.e. $|\delta|$, we sample a value uniformly at random
    from $(0, 1] \cap \Sigma$.


We simulate this system with $q=30$. Figure~\ref{fig:NE_arrival}
demonstrates that the system can still evolve to a state close to
the navigable small-world NE in a few hundred steps, e.g., the
strategies of 80.5\% players fall in the interval $[1.8, 2.4]$, and
the median of the strategies is the navigable NE strategy of $2.1$.
We test $50$ random starting profiles, and take snapshots of the
strategy profiles at the time step $t=500$. On average, the
strategies of 79.8\% players in the snapshots fall in the interval
$[1.8, 2.4]$.
\begin{figure*}[t]
\begin{minipage}[t]{0.48\columnwidth}
    \centerline{\includegraphics[width=1\textwidth, height =0.8\textwidth]{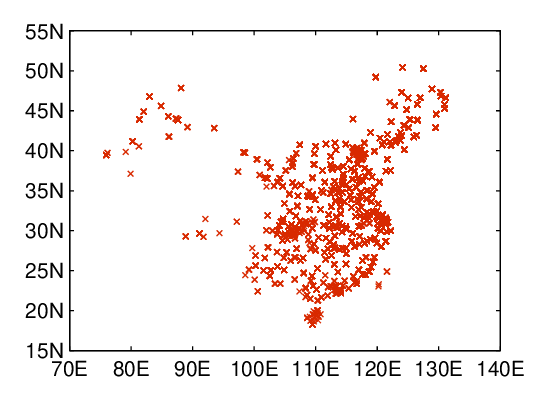}}
    \caption{The hometown location of Renren users.}
    \label{fig:real_grid}
\end{minipage}
\hfill
\begin{minipage}[t]{0.48\columnwidth}
    \centerline{\includegraphics[width=1\textwidth, height =0.8\textwidth]{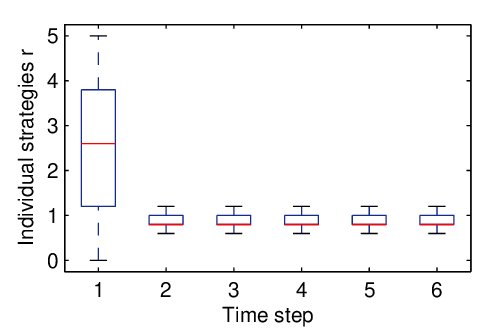}}
    \caption{Simulated network evolution over Renren grid.}
    \label{fig:simulated_evolve}
\end{minipage}
\hfill
\end{figure*}

In summary, our empirical evaluation strongly supports that our payoff function
    considering the balance between link distance and reciprocity
    naturally gives rise to the navigable small-world network.
The convergence to navigable equilibrium
    will happen either when the players know all other players'
    strategies, or only learn their friends' strategies, or only use the
    empirical distance and reciprocity measure.
Once in the navigable equilibrium, the system is very stable and
hard to deviate by any random perturbation.
Furthermore, other equilibria such as the random small world is
    not stable, in that a small perturbation will drive the system back
    to the navigable small-world network.


\subsection{DRB Game under Real Population Distribution}\label{sec:real_eval}
Recall that real population is not evenly distributed
geographically as in the Kleinberg's model. So we want to examine
if our game could lead to an overall connection preference $r$ similar to the
empirical one in the real network. To do so, we examine our game with the non-uniform geographic distribution of people in the following two real networks.
We also introduce
heterogeneity in players' tradeoff functions with $\alpha_u$ taken from a uniform distribution on $[0.1, 10]$.

\para{Renren Network.}We sample 10K users at random from Renren network, and we
construct a real grid through mapping the hometown listed in users' profiles to (longitude, latitude)
coordinates, as shown in Figure~\ref{fig:real_grid}.
To examine the convergence of navigable small-world NE in this
scenario, we simulate the DRB game as follows.
At time step $0$, each player chooses an initial strategy uniformly at random
from the interval $[0, 5]\cap \Sigma $. At each time step $t \geq 1$, every player picks the
best strategy based on the strategies of others in the previous
step: $r_u^t=\argmax_{r_u \in \Sigma\cap [0,5]} \pi
(r_u,\mathbf{r}^{t-1}_{-u}), \forall u, \forall t
>1.$

Figure~\ref{fig:simulated_evolve} shows that in a few steps, the system reaches a NE,
where individual users adopt their respective equilibrium strategies. In the NE,
the mean of the strategies is $0.85$ and the strategies of 88.1\% users fall in the interval [0.7,1.1].
So the overall connection preference of users in the simulated game is very close to
the empirical value of $0.9$ shown in Figure~\ref{fig:link_prob}.

\begin{figure*}[t]
\begin{minipage}[t]{0.33\columnwidth}
    \centerline{\includegraphics[width=1\textwidth, height =0.8\textwidth]{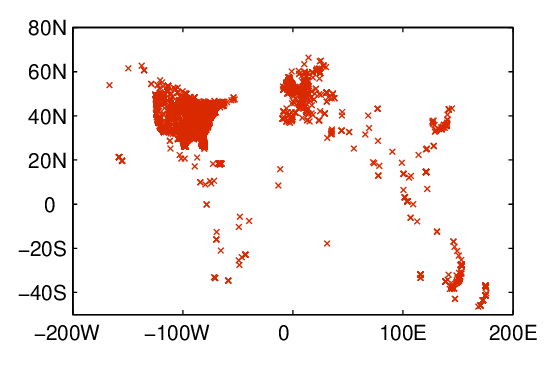}}
    \caption{The hometown location of LiveJournal users.}
    \label{fig:LJ_real_grid}
\end{minipage}
\hfill
\begin{minipage}[t]{0.32\columnwidth}
    \centerline{\includegraphics[width=1\textwidth, height =0.8\textwidth]{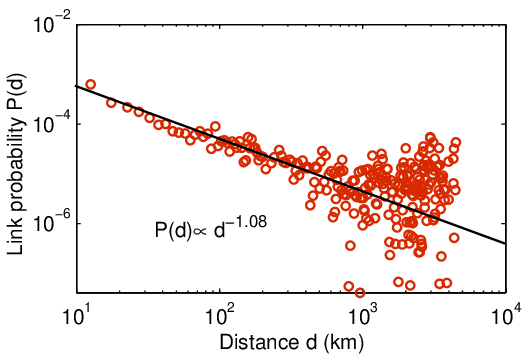}}
    \caption{Friendship probability vs. distance in LiveJournal.}
    \label{fig:LJ_link_prob}
\end{minipage}
\hfill
\begin{minipage}[t]{0.32\columnwidth}
    \centerline{\includegraphics[width=1\textwidth, height =0.8\textwidth]{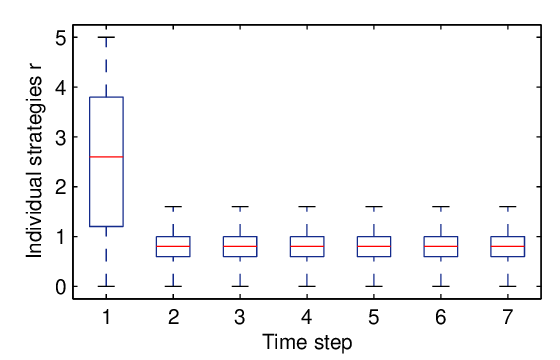}}
    \caption{Simulated network evolution over LiveJournal network.}
    \label{fig:LJ_evolve}
\end{minipage}
\end{figure*}

\para{LiveJournal Network.}To evaluate across-dataset generalization, we also examine our DRB game in the LiveJournal social network.
LiveJournal is a community of bloggers with over 39 million registered users worldwide as the end of 2012.
Each user provides a personal profile, including home location, personal interests
and a list of other bloggers considered as friends. We crawl the profiles of 527,769
LiveJournal users used in the study~\cite{Liben-Nowell}.
Given the 224,155 users providing city
information, we successfully obtained a meaningful geographic
location for only 197,504 users, as shown in Figure~\ref{fig:LJ_real_grid}.
To get the empirical connection preference of these LiveJournal users, we compute the friendship probability $p(d)$ for any given distance $d$
by the proportion of friendships among all
pairs $(u, v)$ with distance $d$. Figure~\ref{fig:LJ_link_prob} shows the relationship
between friendship probability and geographic
distance, which shows that the real connection preference of users is around $1.08$.
These results demonstrates that our game does generalize across online social networks.

\section{Discussion and Future Work}
Our paper is a contribution to the literature on navigability and also on network formation games. There exists a plethora of results relating to network formation
games in economics, as well as in computer science~\cite{Jackson}. Existing games typically use pure-link-based strategy, and it is difficult for an individual to estimate the potential likelihood of forming reciprocal links  to other users (i.e., the introduced notion of reciprocity). These games lead
to mostly trivial equilibria such as cliques or stars. Different from prior games, our game uses the link probability functions as the strategies, and an individual is able to estimate reciprocity by learning the connection preferences of others. This difference in modeling methodology is substantive since it gives rise to the non-trivial structure of navigable small world networks.
Also, most network formation games examine the strategically stable networks in static and non-perturbed settings.
By contrast, our game examines the stability of equilibrium networks under the best response
dynamics with the perturbation introduced. Dynamics could help select among different equilibria of the static
game, the results in this paper illustrate this potential very well.

In the paper, we use a $k$-dimensional grid ($k\in \mathbb{N}\setminus\{0\}$) to be consistent with Kleinberg's small-world model,
where each grid location contains a single node (a total of $n^k$ nodes). Let $B_l(u)$ denote the number of nodes within
distance $l (l > 0)$ from a node $u$. In fact, the key spatial property required in our analysis is $B_l(u) = \Theta(l^D), \forall u\in V, \forall D > 0$,
where $D$ is actually the fractional dimension proposed by Liben-Nowell et al.~\cite{Liben-Nowell}.
So most of our results can be easily extended to a more general space that can be described the fractional dimension, where the grid
is only the special case of integer dimension. The results in Section~\ref{sec:real_eval} actually provided empirical demonstration on Renren and LiveJournal latitude-longitude space that have a fractional dimension close to one.
Similarly, we can also allow multiple nodes to be in the same location as long as the spatial property of $B_l(u) = \Theta(l^D), \forall u\in V$ holds for $l > 0$, i.e., the space could still be described by the fractional dimension.
Given $p=0$ in our setting $K(n, k, p, q, {\bf r})$, each node has undirected edges connecting to all other nodes in the same location as local contacts.
    As we have discussed in Section \ref{sec:DRB-payoff}, we do not consider these local contacts in our game.

The population size and the payoff
obtained at the critical value are sufficiently large to allow us to ignore stochastic effects.
In our game, the environment state consists of (i) the geographical distribution of users, which remains stable over time
given the large population size;
and (ii) the tradeoff factor $\alpha_u$ ($\alpha_u > 0$) for any user $u$, which has no influence on the strategy choice over time, as implied by lemma~\ref{lem:neighbors}.
Specifically, a node $u$ chooses $r_u=k$ once it has a nearby node $w$ satisfying $r_w \geq k$
or $r_w= \max_{v¡ÊV}r_v$, irrespective of the tradeoff factor every node chooses (including the $u$ itself).

Our study opens many possible directions of future work.
For example, one may provide a theoretical analysis of the DRB game on the
	non-uniform population distributions, which has been empirically validated by our
	experiments on the Renren and LiveJournal datasets.
Another direction is to integrate prior studies on human mobility model to provide
   a more complete picture of the underlying mechanisms for navigable
   small-world networks. For example, one could use move-and-forget mobility model~\cite{haintreau08} to generate link probability functions of power law form, and adopt our game-theoretic approach to drive individuals to choose the critical one enabling
navigability to arise.
It is also interesting to investigate the existence of other forms of utility functions,
since given the complex human behavior in the real world, there might be more behavioral factors leading to the actual real small world.
We wish our study could encourage more empirical and theoretical studies on the relationship
	between reciprocity, distance, and navigability, and perhaps uncover the underlying human
	behavior model that integrates these factors together.


%


\bibliographystyle{ACM-Reference-Format-Journals}
\bibliography{yang}

\clearpage
\appendix
\numberwithin{equation}{section}
\makeatletter
\newcommand{\section@cntformat}{Appendix \thesection:\ }
\makeatother

\begin{table}[htb]\caption{Notation}
\centering
\begin{small}
\begin{tabular}{|c|c|c|c|c|}
  \hline
  $k, n$ & Dimension and edge length of a grid $\underbrace{n\times n \times \ldots \times n}_k$  \\\hline
  $n_D$ & Diameter of grid, $n_D = k\lfloor n/2\rfloor$  \\\hline
  $V$ &  Set of players   \\\hline
  $p, q$ & Number of local and long-range contacts  \\\hline
  $d_M(u,v)$ &  Manhattan
distance between players $u$ and $v$   \\\hline
  $r_u$ & Connection preference of player $u$  \\\hline
    $\alpha_u$ & Constant exponent for player $u$'s distance-reciprocity tradeoff, $(\alpha_u>0)$   \\\hline
 $c(r_u)$ &  Normalization constant $c(r_u)=\sum_{\forall v \neq
u}d_M(u, v)^{-r_u}$	  \\\hline
  $p_u(v,r_u)$ & Probability that $u$ connects $v$ under $r_u$, $p_u(v,r_u) =d_M(u, v)^{-r_u}/c(r_u)$ \\\hline
  $\bf r$ & Vector of $r_u$ values on all players (strategy profile)\\\hline
  $\pi_u(r_u, \mathbf{r_{-u}})$ & Player $u$'s payoff given the strategy profile $\bf r$ \\\hline
   $D(r_u)$ & Average link distance of $u$, $D(r_u)=\sum_{\forall v \neq u}{p_u(v, r_u)d_M(u,v)}$ \\\hline
   $P_u(r_u,{\bf r}_{-u})$ & Reciprocity of $u$, $P_u(r_u,{\bf r}_{-u})=\sum_{\forall v \neq u}{p_u(v, r_u)p_v(u, r_v)}$ \\ \hline
   $\gamma$ & Granularity of connection preference, strategy set $\Sigma=\{0, \gamma, 2\gamma, 3\gamma, \ldots, \}$ \\ \hline
   $b_u(j)$ & The number of players at grid distance $j$ from $u$ \\ \hline
   $\xi^-_k$ & Constant making $b_u(j) \geq \xi^-_k j^{k-1}$ for $1\le j\leq \lfloor n/2 \rfloor$ \\ \hline
   $\xi^+_k$ & Constant making $b_u(j) \leq \xi^+_k j^{k-1}$ for  $j \leq n_D$ \\ \hline
   $\varepsilon$ & Deviation from strategy $k$, $\varepsilon = k-r_u$ \\
  \hline
\end{tabular}
\end{small}
\label{tab:notation}
\end{table}

\section{Commonly used results on the Kleinberg's small world and the DRB game}
In all proofs in the appendix, for a given node $u\in V$,
    we denote $D(r_u)=\sum_{\forall v \neq u}{p_u(v, r_u)d_M(u,v)}$
    as its average grid distance of its long range contacts (simply
    referred to as the {\em link distance}), and
    $P_u(r_u,{\bf r}_{-u})$ $=$ $\sum_{\forall v \neq u} $
        ${p_u(v, r_u)p_v(u, r_v)}$ as
    its {\em reciprocity}.
When ${\bf r}_{-u} \equiv s$, we simply use $P(r_u, s)$ to denote
    $P_u(r_u, {\bf r}_{-u} \equiv s)$. Moreover, for any $A \subseteq V$,
let $P_{u,A}({\bf r}) = \sum_{v\in A} p_u(v, r_u) p_v(u,r_v)$ be the reciprocity
    $u$ obtained from subset $A$.
We denote $c(r_u)$ $=$ $\sum_{\forall v \neq u}{d_M(u,v)^{-r_u}}$ as
    $u$'s normalized coefficient.
The subscript $u$ in $D(r_u)$, $P(r_u, s)$ and $c(r_u)$ is omitted because
    their values are the same for all $u\in V$.

Let $n_D$ be the longest grid distance among nodes in
    $K(n, k, p, q, {\bf r})$.
We have that $n_D = k\lfloor n/2\rfloor$. We denote $b_u(j)$ as the
number of players at grid distance $j$ from $u$.
We can find two constants $\xi^-_k$ and $\xi^+_k$ only depending on
the dimension $k$, so that $\xi^-_k j^{k-1} \leq b_u(j)\leq \xi^+_k
j^{k-1}$ for $1\le j\leq \lfloor n/2 \rfloor$ and $1 \le b_u(j) \leq
\xi^+_k j^{k-1}$ for $\lfloor n/2 \rfloor <j\leq n_D$.\footnote{The
    exact values of $\xi^-_k$ and $\xi^+_k$ can be derived by the
    combinatorial problem of counting the number of ways to choose
    $k$ non-negative integers such that they sum to a given positive integer
    $j$.}
Note that the payoff function for the DRB
    game is indifferent of parameters $p$ and $q$ of the network, so we
    treat $p=q=1$ for our convenience in the analysis.

Recall that we assume that each $r_u$ is taken from a discrete set
    $\Sigma=\{0, \gamma, 2\gamma, 3\gamma, \ldots, \}$, where
    $\gamma$ represents the granularity of connection preference and is
    in the form of $1/g$ for some positive integer $g \ge 2$.
Using discrete strategy set avoids nuances in continuous strategy space and
    is also reasonable in practice since people are unlikely to
    make infinitesimal changes. Henceforth, for any $r_u \neq k$, we have $|k-r_u| \geq \gamma$.
    The notation
commonly used in the paper is described in Table~\ref{tab:notation}.

We first show the following two lemmas, which will be used in the most of theorems.
\begin{lemma} \label{lem:normlizationbound}
In the $k$-dimensional grid $K(n, k, p, q, {\bf r})$, for a given node $u\in V$ with a strategy of $r_u$,
the normalized coefficient $c(r_u)$ has
the following bounds:
\begin{small}
   \begin{subnumcases}{}
   \frac{\xi^-_k}{2^{k+1}k}n^{k-r_u} \leq c(r_u) \leq \xi^+_k k^{k-r_u}n^{k-r_u}   &\mbox{if $r_u < k$,} \label{enq:pr2a}\\
    \frac{\xi^-_k\ln n}{2} \leq c(r_u)\leq \xi^+_k\ln(2kn) &\mbox{if $r_u = k$,} \label{enq:pr2b}\\
    \xi^-_k \leq c(r_u) \leq \xi^+_k(1+1/\gamma) &\mbox{if $r_u > k$}. \label{enq:pr2c}
   \end{subnumcases}\label{enq:pr2}
\end{small}
\end{lemma}
\begin{proof}
In the case of $r_u < k$, we write $\varepsilon = k-r_u$ $(\gamma \le \varepsilon \le k)$.
The coefficient $c(r_u)$ can be bounded as:
\begin{small}
\begin{align}
c(r_u) = & \sum_{\forall v \neq u}{d_M(u,v)^{-r_u}} \geq
\sum_{j=1}^{n/2}b_u(j)j^{-r_u} \geq
\xi^-_k\sum_{j=1}^{n/2}j^{\varepsilon-1}
    \nonumber \\
& \geq \xi^-_k\int_{1}^{n/2}x^{\varepsilon-1}dx  \geq
\frac{\xi^-_k}{\varepsilon}\left(\frac{n}{2}\right)^\varepsilon-\frac{\xi^-_k}{\varepsilon} \nonumber
\ge
\frac{\xi^-_k}{2\varepsilon}\left(\frac{n}{2}\right)^\varepsilon \nonumber
\end{align}
\end{small}
The last inequality above relies on a loose relaxation of
     $\frac{1}{2}\left(\frac{n}{2}\right)^\varepsilon \ge 1$,
     which is guaranteed for all $n\ge 2^{1+1/\gamma}$
     since $\varepsilon \ge \gamma$. Note that $\varepsilon <k$, so we have:
\begin{small}
\begin{align}
c(r_u)
\ge
\frac{\xi^-_k}{2^{1+\varepsilon}\varepsilon}n^\varepsilon \ge \frac{\xi^-_k}{2^{1+k}k}n^\varepsilon .\nonumber
\label{enq:cr1}
\end{align}
\end{small}

The upper bound of coefficient $c(r_u)$ can be given as:
\begin{small}
\begin{align}
c(r_u) = & \sum_{\forall v \neq u}{d_M(u,v)^{-r_u}} =
\sum_{j=1}^{n_{D}}b_u(j)j^{-r_u} \leq
\xi^+_k\sum_{j=1}^{n_D}j^{\varepsilon-1} \nonumber \\& \leq
   \begin{cases}
   1+ \xi^+_k\int_{1}^{n_D}j^{\varepsilon-1}dx \leq 1+ (kn/2)^{\varepsilon} &\mbox{if $ \varepsilon < 1$,}  \nonumber \\
   \xi^+_k\int_{j=1}^{n_D+1}j^{\varepsilon-1}dx \leq \xi^+_k(kn/2+1)^{\varepsilon}  &\mbox{if $\varepsilon \geq 1$,}  \nonumber
   \end{cases} \nonumber \\ & \leq \xi^+_k k^{\varepsilon}n^{\varepsilon}. \nonumber
\end{align}
\end{small}
The last inequality above relies on a loose relaxation of
     $\frac{kn}{2} \ge 1$, which is guaranteed for all $n\ge 2$
     since $k \ge 1$.

We now turn to the case of $r_u = k$. The upper bound of normalization coefficient $c(k)$ can be given
as
\begin{small}
\begin{equation}
c(k) = \sum_{\forall v \neq u}{d_M(u,v)^{-k}} =
\sum_{j=1}^{n_D}b_u(j)j^{-k} \leq \xi^+_k\sum_{j=1}^{n_D}\frac{1}{j} \leq
\xi^+_k\ln(2kn) \nonumber, \label{enq:ck}
\end{equation}
\end{small}
and its lower bound is
\begin{small}
\begin{equation} \label{enq:cklower}
c(k) \geq \xi^-_k\sum_{j=1}^{n/2}j^{-1} \geq
\xi^-_k\int_{1}^{n/2}x^{-1}dx \geq \xi^-_k(\ln n-\ln2) \ge
\frac{\xi^-_k\ln n}{2}\nonumber.
\end{equation}
\end{small}
where the last inequality is true when $n \ge e^4$.

We finally consider the the case of $r_u>k$, it is easy to get that
\begin{small}
\begin{equation}
c(r_u) = \sum_{\forall v \neq u}{d_M(u,v)^{-r_u}} \geq
\sum_{j=1}^{n/2}b_u(j)j^{-r} \geq b_u(1) \geq \xi^-_k \nonumber,
\label{enq:cbr}
\end{equation}
\end{small}
and its upper bound is given as:
\begin{small}
\begin{align}
c(r_u)= \sum_{\forall v \neq u}{d_M(u,v)^{-r_u}} =
\sum_{j=1}^{n_D}b_u(j)j^{-r_u} \leq  \xi^+_k \sum_{j=1}^{n_D}j^{k-1}j^{-r_u} \leq \xi^+_k\sum_{j=1}^{n_D}j^{-(r_u-k)-1}. \nonumber
\end{align}
\end{small}

Note $r_u-k\geq \gamma$, we have:
\begin{small}
\begin{align}
c(r_u) \leq \xi^+_k\sum_{j=1}^{n_D}j^{-(r_u-k)-1}  \leq \xi^+_k\sum_{j=1}^{n_D}j^{-\gamma-1}  \leq
\xi^+_k(1+\int_{1}^{n_D+1}x^{-\gamma-1}dx)
\leq
\xi^+_k\left(1+\frac{1}{\gamma}\right). \nonumber
\end{align}
\end{small}

\end{proof}

\begin{lemma} \label{lem:distancebound}
In the $k$-dimensional grid $K(n, k, p, q, {\bf r})$, for a given node $u\in V$ with a strategy of $r_u$,
the average distance of its long-range contacts $D(r_u)$ has
the following bounds:
\begin{small}
   \begin{subnumcases}{}
    D(r_u) \leq  \frac{\xi^+_k k^{1+k}}{ c(r_u)} n^{1+k-r_u} &\mbox{if $r_u < k$,}\label{enq:distancebounda}\\
     \frac{\xi_k^- n}{2c(k)} \leq D(r_u)\leq \frac{\xi_k^+ n}{c(k)} &\mbox{if $r_u = k$,} \label{enq:distanceboundb}\\
    D(r_u) \leq \frac{\xi^+_kk}{2\gamma c(r_u)}n^{1-\gamma} &\mbox{if $k< r_u < k+1$,} \label{enq:distanceboundc}\\
    D(r_u) \leq \frac{\xi^+_k}{c(r_u)}\ln(2kn). &\mbox{if $r_u \geq k+1$} \label{enq:distanceboundd}.
   \end{subnumcases}\label{enq:distancebound}
\end{small}
\end{lemma}
\begin{proof}
When $r_u < k$, we write $\varepsilon = k-r_u$ $(\gamma \le \varepsilon \le k)$ and get
the upper bound for the link distance
\begin{small}
\begin{align}
D(r_u) &= \frac{\sum_{j=1}^{n_D}b_u(j)\cdot j^{-r_u} \cdot j}{c(r_u)} \leq
\frac{\xi^+_k\int_{1}^{n_D+1}x^{\varepsilon}dx}{c(r_u)} \leq \frac{\xi^+_k (n_D+1)^{1+\varepsilon}}{(1+\varepsilon)
c(r_u)}
    \le \frac{\xi^+_k (kn)^{1+\varepsilon}}{c(r_u)}\le \frac{\xi^+_k k^{1+k}}{ c(r_u)} n^{1+\varepsilon}.\nonumber
\label{enq:dr}
\end{align}
\end{small}

We now turn to the case of $r_u = k$. The upper bound of link distance $D(k)$ can be given
as
\begin{small}
\begin{equation}
D(r_u=k) = \frac{\sum_{j=1}^{n_D}b_u(j)\cdot j^{-k} \cdot j}{c(r_u)}
    \leq  \frac{\xi_k^+ n}{c(k)},\nonumber
\label{enq:dklower}
\end{equation}
\end{small}
and its lower bound is
\begin{small}
\begin{equation}
D(r_u) \geq \frac{\sum_{j=1}^{n/2}b_u(j)\cdot j^{-k} \cdot j}{c(k)}
\geq \frac{\xi_k^- n}{2c(k)}.\nonumber \label{enq:dk}
\end{equation}
\end{small}

We finally consider the case of $r_u>k$.
We write $\varepsilon = r_u-k
(\varepsilon \ge \gamma)$, and the bound for the link distance is:
\begin{small}
\begin{equation}
\begin{aligned}
D(r_u) & = \frac{\sum_{j=1}^{n_D}b_u(j)\cdot j^{-r_u} \cdot
j}{c(r_u)} \leq
\xi^+_k\sum_{j=1}^{n_D}{\frac{j^{-\varepsilon}}{c(r_u)}} \leq
\xi^+_k\frac{1+\int_{1}^{n_D}x^{-\varepsilon}dx}{c(r_u)} \nonumber
\label{enq:dbr}
\end{aligned}
\end{equation}
\end{small}
In the case of $\varepsilon < 1$,
\begin{small}
\begin{align}
D(r_u) & \leq
\xi^+_k\frac{1+\int_{1}^{n_D}x^{-\varepsilon}dx}{c(r_u)} \leq
   \frac{\xi^+_k}{(1-\varepsilon)c(r_u)}(kn/2)^{1-\varepsilon} 
   \leq \frac{\xi^+_kk}{2\gamma c(r_u)}n^{1-\varepsilon} \leq \frac{\xi^+_kk}{2\gamma c(r_u)}n^{1-\gamma}.\nonumber
\label{enq:dbr}
\end{align}
\end{small}
otherwise,
\begin{small}
\begin{equation}
D(r_u) \leq
\xi^+_k\frac{1+\int_{1}^{n_D}x^{-\varepsilon}dx}{c(r_u)} \leq \frac{\xi^+_k}{c(r_u)}\ln(2kn).\nonumber
\label{enq:dbr}
\end{equation}
\end{small}
\end{proof}

\begin{lemma} \label{lem:loworder}
In the $k$-dimensional DRB game, there exists a constant $\kappa$
    (only depending on model constants $k$ and $\gamma$),
    for sufficiently large $n$ (in particular $n \ge \max(e^4, 2k)$),
    the following statement holds:
    for any strategy profile $\bf r$,
    any node $u$ with $r_u \ne k$ and $\alpha_u>0$,
    $\pi_u(r_u, {\bf r}_{-u}) \le \kappa n^{\alpha_u-\min\{1, \alpha_u\}\gamma}$.
\end{lemma}
\begin{proof}
We introduce some notations first.
Given the strategy profile $\bf r$ and a node $u$ with $r_u \ne k$,
    we partition the rest nodes $V\setminus \{u\}$ into three sets:
    $V_{< k} = \{v\in V\setminus \{u\} \mid r_v < k\}$,
    $V_{> k} = \{v\in V \setminus \{u\} \mid r_v > k\}$,
    $V_{ = k} = \{v\in V \setminus \{u\}\mid r_v = k\}$.
Then we have
\begin{small}
\begin{equation}
\pi_u(\mathbf{r}) = D(r_u)^{\alpha_u}\left( P_{u,V_{<k}}({\bf r}) +
    P_{u,V_{>k}}({\bf r}) + P_{u,V_{=k}}({\bf r}) \right).
\end{equation}
\end{small}

We now consider the case of $r_u<k$ and $r_u>k$ separately.

\para{Payoff of $r_u<k$.}
Let $\varepsilon=k-r_u$ ($\gamma \le \varepsilon \le k$).
We first consider the average grid distance to long-range contacts in this case.
Based on the bound on $D(r_u)$ and $c(r_u)$ given by inequalities \eqref{enq:distancebounda} and \eqref{enq:pr2a}, we get:
\begin{small}
\begin{align}
D(r_u)^{\alpha_u} = \frac{\xi^+_k k^{1+k}}{ c(r_u)} n^{1+\varepsilon}
= \frac{\xi^+_k k^{1+k}}{ \frac{\xi^-_k}{2^{k+1}k}n^{\varepsilon}} n^{1+\varepsilon}
= \frac{2^{k+1}\xi^+_k k^{2+k}}{\xi^-_k} n \label{eqn:distsk}
\end{align}
\end{small}

We now examine the reciprocity. We first consider the reciprocity
player $u$ obtains from the players in $V_{< k}$.
We have $c(r_v) \ge c(k-\gamma) $ for $\forall v \in V_{ < k}$, since
    $r_v \le k-\gamma$.
Then we have:
\begin{small}
\begin{align}
P_{u, V_{ < k}}({\bf r}) &  =  \sum_{v \in V_{ < k}}
    \frac{d_M(u,v)^{-r_u-r_v}}{c(r_u)c(r_v)}
    \le \sum_{v \in V_{ < k}}\frac{d_M(u,v)^{-r_u}}{c(r_u)c(k-\gamma)}  \leq \frac{\sum_{\forall v \ne u}
d_M(u,v)^{-r_u}}{c(r_u)c(k-\gamma)} = \frac{1}{c(k-\gamma)}. \nonumber
\end{align}
\end{small}

Combining with the
inequalities \eqref{eqn:distsk} and \eqref{enq:pr2a}, we get:
\begin{small}
\begin{equation}
D(r_u)^{\alpha_u} P_{u,V_{<k}}({\bf r}) \le    \left(\frac{2^{k+1}\xi^+_k k^{2+k}}{\xi^-_k} n\right)^{\alpha_u} \frac{1}{\frac{\xi^-_k}{2^{k+1}k}n^{\gamma}} \le
\frac{(\xi^+_k)^{\alpha_u}2^{(k+1)(\alpha_u+1)} k^{(k+2)\alpha_u+1}}{(\xi^-_k)^{\alpha_u+1}}n^{\alpha_u-\gamma}.\label{enq:urr}
\end{equation}
\end{small}

Next we examine the reciprocity that player $u$ obtains from
the players in $V_{>k}$.
Note that for all $v\in V_{>k}$, $r_v \ge k + \gamma$.
Using the bound on $c(r_v)$ given by inequality \eqref{enq:pr2c}, we have:
\begin{small}
\begin{equation}
\begin{aligned}
& P_{u, V_{>k}}({\bf r})
 =  \sum_{v \in V_{ < k}}
    \frac{d_M(u,v)^{-r_u-r_v}}{c(r_u)c(r_v)}  
    \leq \frac{\sum_{j=1}^{n_D}{b_u(j)\cdot j^{-r_u} \cdot
j^{-k-\gamma}}}{\xi^-_k c(r_u)} =
\frac{\xi^+_k\sum_{j=1}^{n_D}j^{-1-r_u-\gamma}}{\xi^-_kc(r_u)}\\&
\leq
\frac{\xi^+_k(1+\int_{1}^{n_D}x^{-1-r_u-\gamma}dx)}{\xi^-_kc(r_u)} \leq
\frac{\xi^+_k(1+r_u+\gamma)}{\xi^-_k(r_u+\gamma) c(r_u)} \leq
\frac{\xi^+_k(k+1)}{\xi^-_k\gamma c(r_u)}. \nonumber
\end{aligned}
\end{equation}
\end{small}

Based on the bound on $D(r_u)$ and $c(r_u)$ given by inequalities \eqref{eqn:distsk} and \eqref{enq:pr2a}, we get:
\begin{small}
\begin{equation}
\begin{aligned}
D(r_u)^{\alpha_u} P_{u,V_{>k}}({\bf r}) & \leq  \left(\frac{2^{k+1}\xi^+_k k^{2+k}}{\xi^-_k} n\right)^{\alpha_u} \cdot \frac{\xi^+_k(k+1)}{\xi^-_k\gamma \frac{\xi^-_k}{2^{k+1}k}n^{\varepsilon}}
\\& \leq \frac{(\xi^+_k)^{\alpha_u+1}
2^{(k+1)(\alpha_u+1)}k^{(k+2)\alpha_u+1}(k+1)}{(\xi^-_k)^{\alpha_u+2}\gamma}n^{\alpha_u-\varepsilon} \\& 
\leq \frac{(\xi^+_k)^{\alpha_u+1} 2^{(k+1)(\alpha_u+1)+1}k^{(k+2)\alpha_u+2}}{(\xi^-_k)^{\alpha_u+2}\gamma}n^{\alpha_u-\gamma}.
\label{enq:urbr}
\end{aligned}
\end{equation}
\end{small}

We now examine the payoff of player $u$ from players in $V_{=k}$.
In this case, the upper bound for the reciprocity is:
\begin{small}
\begin{equation}
P(r_u,k) = \frac{\sum_{j=1}^{n_D}b_u(j)\cdot j^{-r_u} \cdot
j^{-k}}{c(r_u)c(k)} \leq \xi^+_k\sum_{j=1}^{n_D}\frac{
j^{\varepsilon-1-k}}{c(r_u)c(k)}. \label{enq:pr} \nonumber
\end{equation}
\end{small}

Notice that $\varepsilon \le k$, we have:
\begin{small}
\begin{align}
P(r_u,k) \leq \frac{\xi^+_k
}{c(r_u)c(s)}\left(1+\int_{j=1}^{n_D}j^{\varepsilon-1-s}\right) 
\leq
   \begin{cases}
   \frac{\xi^+_k(1+k-\varepsilon)}{(k-\varepsilon)c(r_u)c(k)} \leq \frac{ 2k\xi^+_k}{
   \gamma c(r_u)c(k)} &\mbox{if $ \varepsilon < k$,} \\
   \frac{2\xi^+_k \ln(2kn)}{c(r_u)c(k)} &\mbox{if $\varepsilon = k$,}  \nonumber
   \end{cases}
\end{align}
\end{small}
The inequalities in the cases above
    use the facts $\gamma \le \varepsilon \le k$ and $k-\varepsilon \ge \gamma$ when
    $\varepsilon < k$.

Combining the the above bounds on reciprocity with bounds given by
inequalities \eqref{eqn:distsk}, \eqref{enq:pr2a} and \eqref{enq:pr2b},
    we have the payoff of node $u$ getting from $V_{=k}$:
\begin{small}
\begin{align}
& D(r_u)^{\alpha_u} P_{u,V_{=k}}({\bf r}) \leq D(r_u)^{\alpha_u}P(r_u,k) \nonumber \\
& \leq
\begin{cases}
  \left(\frac{2^{k+1}\xi^+_k k^{2+k}}{\xi^-_k} n\right)^{\alpha_u} \cdot \frac{ 2k\xi^+_k}{
   \gamma \frac{\xi^-_k n^{\varepsilon}}{2^{k+1}k} \cdot \frac{\xi^-_k\ln n}{2}} &\mbox{if $ \varepsilon < k$,} \\
    \left(\frac{2^{k+1}\xi^+_k k^{2+k}}{\xi^-_k} n\right)^{\alpha_u} \cdot \frac{2\xi^+_k \ln(2kn)}{\frac{\xi^-_k n^{\varepsilon}}{2^{k+1}k} \cdot \frac{\xi^-_k\ln n}{2}} &\mbox{if $\varepsilon = k$,}  \nonumber
   \end{cases}\\ & \leq
   \begin{cases}
   \frac{(\xi^+_k)^{\alpha_u+1}2^{\alpha_u(k+1)+k+3
   }k^{\alpha_u(k+2)+1}}{\gamma (\xi^-_k)^{\alpha_u+2}}\frac{n^{\alpha_u-\gamma}}{\ln n}
   &\mbox{if $\varepsilon < k$}.\\
   \frac{(\xi^+_k)^{\alpha_u+1}2^{\alpha_u(k+1)+k+3}k^{\alpha_u(k+2)+1}\ln (2kn) }{(\xi^-_k)^{\alpha_u+2} \ln n}n^{\alpha_u-\gamma}
   \le \frac{(\xi^+_k)^{\alpha_u+1}2^{\alpha_u(k+1)+k+4}k^{\alpha_u(k+2)+1}}{(\xi^-_k)^{\alpha_u+2}}n^{\alpha_u-\gamma}
   &\mbox{if $\varepsilon = k$}.\\
   \end{cases}
\label{enq:urk}
\end{align}
\end{small}

The last inequality in the case of $\varepsilon = k$ requires $n\ge 2k$.

Adding up results in Eq.\eqref{enq:urr}, \eqref{enq:urbr}, \eqref{enq:urk},
    we obtain that
    \begin{small}
    \begin{equation}
\begin{aligned}
\pi(r_u, \mathbf{r}_{-u}) & \le \frac{(\xi^+_k)^{\alpha_u}2^{(k+1)(\alpha_u+1)} k^{(k+2)\alpha_u+1}}{(\xi^-_k)^{\alpha_u+1}}n^{\alpha_u-\gamma} + \frac{(\xi^+_k)^{\alpha_u+1} 2^{(k+1)(\alpha_u+1)+1}k^{(k+2)\alpha_u+2}}{(\xi^-_k)^{\alpha_u+2}\gamma}n^{\alpha_u-\gamma} +  \\ & \frac{(\xi^+_k)^{\alpha_u+1}2^{\alpha_u(k+1)+k+4}k^{\alpha_u(k+2)+1}}{(\xi^-_k)^{\alpha_u+2}}n^{\alpha_u-\gamma} \\ & \le \frac{3(\xi^+_k)^{\alpha_u+1}\cdot
2^{\alpha_u(k+1)+k+4}k^{\alpha_u(k+2)+2}}{\gamma (\xi^-_k)^{\alpha_u+2}}n^{\alpha_u-\gamma} \\&
\le
\frac{(\xi^+_k)^{\alpha_u+1}\cdot
2^{\alpha_u(k+1)+k+6}k^{\alpha_u(k+2)+2}}{\gamma (\xi^-_k)^{\alpha_u+2}}n^{\alpha_u-\gamma},
\label{enq:lessk}
\end{aligned}
\end{equation}
\end{small}
when $n\ge \max \{e^4, 2k\}$.

\para{Payoff of $r_u>k$.}
Let $\varepsilon = r_u-k$ ($\varepsilon \ge \gamma$).
For this case, we can relax the reciprocity $P_u(r_u, {\bf r}_{-u})$ to one
    and only consider the upper bound on link distance $D(r_u)$.
Applying bounds given by inequalities \eqref{enq:distanceboundc}, \eqref{enq:distanceboundd} and \eqref{enq:pr2a}, we obtain:
\begin{small}
\begin{align}
\pi(r_u = k+\varepsilon, \mathbf{r}_{-u}) \leq D(r_u)^{\alpha_u} &  \leq
\begin{cases}
   \left(\frac{\xi^+_kk}{2\gamma \xi^-_k}n^{1-\gamma}\right)^{\alpha_u}
   &\mbox{if $\varepsilon < 1$,}\\
   \left(\frac{\xi^+_k}{\xi^-_k}\ln(2kn)\right)^{\alpha_u}  &\mbox{if $\varepsilon \geq 1$.}
\end{cases} \nonumber \\ &  \leq
\begin{cases}
   \left(\frac{\xi^+_kk}{\xi^-_k\gamma}\right)^{\alpha_u} n^{\alpha_u(1-\gamma)}
   &\mbox{if $\varepsilon < 1$,}\\
   \left(\frac{\xi^+_k}{\xi^-_k}\right)^{\alpha_u} \ln(2kn)^{\alpha_u} \le  \left(\frac{\xi^+_k}{\xi^-_k}\right)^{\alpha_u}2 n^{\alpha_u(1-\gamma)}  &\mbox{if $\varepsilon \geq 1$.}
   \end{cases}
   \label{enq_ubr}
\end{align}
\end{small}
The last inequality in the above case of $\varepsilon \geq 1$ holds when
    $n \ge 2k$ and $\gamma \le 1/2$.

Finally, the lemma holds when we combine Eq.\eqref{enq:lessk} and \eqref{enq_ubr}
\end{proof}

\begin{lemma} \label{lem:payoffbound}
In the $k$-dimensional DRB game, for sufficiently large $n$,
the payoff of any node $u \in V$ with $\alpha_u>0$ in the navigable small world $\mathbf{r} \equiv k$
has the following bounds:
\begin{small}
\begin{equation}
\frac{(\xi^-_k)^{\alpha_u+1}}{2^{\alpha_u}(\xi^+_k)^{\alpha_u+2}}\frac{n^{\alpha_u}}{\ln^{\alpha_u+2}(2kn)} \leq \pi(r_u = k, \mathbf{r}_{-u} \equiv k) \leq  \frac{2^{3+\alpha_u}(\xi_k^+)^{\alpha_u+1}}{(\xi_k^-)^{2+\alpha_u}}\cdot\frac{n^{\alpha_u}}{\ln^{2+\alpha_u}n}
.\label{enq:ukbound}
\end{equation}
\end{small}
\end{lemma}
\begin{proof}
We have the lower bound for the reciprocity:
\begin{small}
\begin{equation}
P(r_u,k) \geq \frac{\sum_{j=1}^{n/2}b_u(j)\cdot j^{-2k}}{c^2(k)} \geq \frac{\xi^-_k}{c^2(k)}. \nonumber
\end{equation}
\end{small}
Combining the above inequality with bounds~
\eqref{enq:distanceboundb} and \eqref{enq:pr2b}, we get.
\begin{small}
\begin{equation}
\pi(r_u = k, k) = D(r_u)^{\alpha_u}P(r_u,k) \geq \frac{(\xi_k^-)^{\alpha_u+1} n^{\alpha_u}}{2c^{\alpha_u+2}(k)} =
\frac{(\xi^-_k)^{\alpha_u+1}}{2^{\alpha_u}(\xi^+_k)^{\alpha_u+2}}\frac{n^{\alpha_u}}{\ln^{\alpha_u+2}(2kn)}. \nonumber
\end{equation}
\end{small}
The upper bound on the reciprocity is:
\begin{small}
\begin{align}
P(r_u, k)  = \frac{\sum_{j=1}^{n_D}b_u(j)\cdot j^{-2k}}{c^2(k)} 
\leq \frac{\xi^+_k
}{c^2(k)}\left(1+\int_{j=1}^{n_D}j^{-k-1}\right) \leq \frac{2\xi^+_k }{c^2(k)} \nonumber
\end{align}
\end{small}
Combining the above inequality with bounds~
\eqref{enq:distanceboundb} and \eqref{enq:pr2b}, we get the upper bound on the payoff:
\begin{small}
\begin{equation}
\pi(r_u = k, \mathbf{r}_{-u} \equiv k) = D(r_u)^{\alpha_u}P(r_u,k) \leq \frac{2(\xi_k^+)^{\alpha_u+1} n^{\alpha_u}}{c^{\alpha_u+2}(k)}
 \leq \frac{2^{\alpha_u+3}(\xi_k^+)^{\alpha_u+1}}{(\xi_k^-)^{2+\alpha_u}}\cdot\frac{n^{\alpha_u}}{\ln^{2+\alpha_u}n}.  \nonumber
\label{enq:uklowerlk}
\end{equation}
\end{small}
\end{proof}

\section{Proof of Lemma 3.1}\label{proof3}
\neighbors*
\begin{proof}
For a given node $w$, define the set of nodes with distance of $\delta$
to $w$ as: $N_{w,\delta}= \{u|u\in V  \wedge d_M(u,w) \leq \delta \}.$

In the case of $r_w \geq k$, for any $u \in N_{w, \delta}$, if $u$ chooses the strategy $r_u=k$,
we have:
\begin{small}
\begin{align}
P(r_u, {\bf r_{-u}}) > p_u(w, r_u)p_w(u, r_w) \ge \frac{d_M(u,w)^{-2k}}{c(k)^2} \geq \frac{\delta^{-2k}}{c(k)^2}. \nonumber
\end{align}
\end{small}
Combining the above inequality with the bounds in \eqref{enq:distanceboundb} and \eqref{enq:pr2b}, we get:
\begin{small}
\begin{align}
\pi(r_u = k, \mathbf{r}_{-u}) \ge D(r_u)^{\alpha_u} P(r_u, {\bf r_{-u}})  \ge \left(\frac{\xi_k^- n}{2c(k)}\right)^{\alpha_u} \cdot \frac{\delta^{-2k}}{c(k)^2} \ge \frac{(\xi_k^-)^{\alpha_u}\delta^{-2k}}{2^{\alpha_u}(\xi^+_k)^{\alpha_u+2}}\frac{n^{\alpha_u}}{\ln^{\alpha_u+2}(2kn)}.\label{equbound}
\end{align}
\end{small}
However, if node $u$ chooses $r_u \neq k$, by Lemma~\ref{lem:loworder}
we know that there is a constant $\kappa$ such that
for all sufficiently large $n$, $\pi(r_u, \mathbf{r}_{-u})\le \kappa n^{\alpha_u-\min\{1,\alpha_u\}\gamma}$.
We see that the lower bound~\eqref{equbound} for $r_u=k$ is in strictly higher order in $n$ than the upper bound of $r_u\neq k$,
thus there exists $n_0\in \mathbb{N}$ ($n_0$ may depend on $\delta$), such that
	for all $n\ge n_0$,
	$r_u=k$ is the unique best response to $\bf r_{-u}$ for any $u \in N_{w,\delta}$.

In the case of $r_w = \max_{v\in V} r_v$, if $r_w \geq k$, from the above analysis
we know that $r_u=k$ is the unique best response to $\bf r_{-u}$
for any $u \in N_{w,\delta}$.

Otherwise, given $r_w < k$, we know $V = V_{<k}$.
In this case, we further partition the nodes $V_{<k}$ into two sets:
$V_{>0} = \{v\in V \mid k> r_v > 0\}$ and $V_{= 0} = \{v\in V \mid r_v = 0\}$.
So we know that:
\begin{small}
\begin{equation}
\pi_u(\mathbf{r}) = D(r_u)^{\alpha_u}\left( P_{u,V_{>0}}({\bf r}) +
    P_{u,V_{=0}}({\bf r}) \right).
    \label{enq:payoffsum}
\end{equation}
\end{small}
Let $r_{min}$ be the minimum value
among the strategies of users in the set of $V = V_{>0}$.
Clearly, $\gamma \leq r_{min} \leq r_w < k$.

\para{Payoff of $r_u < k$.}For any node $u \in N_{w, \beta}$, if it chooses $r_u < k$, let $\varepsilon = k-r_u$
	$(\gamma \le \varepsilon \le k)$. Notice that $c(v) \geq c(r_w)$ for any
node $v\in V \setminus \{u\}$, so we have:
\begin{small}
\begin{align}
P_{u, V_{>0}}({\bf r}) &  =  \sum_{v \in V_{>0}}
    \frac{d_M(u,v)^{-r_u-r_{v}}}{c(r_u)c(r_v)}  
    \le \sum_{v \in V\setminus \{u\}}\frac{d_M(u,v)^{-r_u-r_{min}}}{c(r_u)c(r_w)} 
    \leq \frac{\sum_{j=1}^{n_D} \xi_k^+ j^{\varepsilon-1-r_{min}}}{c(r_u)c(r_w)}.\nonumber
\end{align}
\end{small}
If $r_{min} + 1\leq \varepsilon$, we have:
\begin{small}
\begin{align}
P_{u, V_{>0}}({\bf r}) & \leq \xi^+_k\int_{j=1}^{n_D+1}\frac{
j^{\varepsilon-r_{min}-1}}{c(r_u)c(r_w)}\leq \frac{\xi_k^+
(n_D+1)^{\varepsilon-r_{min}}}{(\varepsilon-r_{min})c(r_u)c(r_w)} 
\leq \frac{\xi^+_k (kn)^{\varepsilon-r_{min}}}{c(r_u)c(r_w)}. \nonumber
\end{align}
\end{small}
If $r_{min} + 1 > \varepsilon$, we have:
\begin{small}
\begin{align}
P_{u, V_{>0}}({\bf r}) \leq \frac{\xi^+_k
}{c(r_u)c(r_w)}\left(1+\int_{j=1}^{n_D}j^{\varepsilon-1-r_{min}}\right) 
\leq
   \begin{cases}
   \frac{\xi^+_k((kn)^{\varepsilon-r_{min}}+\varepsilon-r_{min}-1)}{(\varepsilon-r_{min})c(r_u)c(r_w)} &\mbox{if $r_{min} \neq \varepsilon$,}\\
   \frac{\xi^+_k \ln(kn)}{c(r_u)c(r_w)} &\mbox{if $r_{min} = \varepsilon$.}
   \end{cases} \nonumber
\end{align}
\end{small}
Combining the above bounds on reciprocity with bounds in inequalities \eqref{eqn:distsk} and \eqref{enq:pr2a}, we have the payoff of node $u$ gets from
$ V_{>0}$:
\begin{small}
\begin{align}
D(r_u)^{\alpha_u} P_{u, V_{>0}}({\bf r}) & \leq
\begin{cases}
   D(r_u)^{\alpha_u}  \cdot \frac{\xi^+_k(kn)^{\varepsilon-r_{min}}}{\gamma c(r_u)c(r_w)}, \mbox{if $\varepsilon > r_{min}$,}\\
   D(r_u)^{\alpha_u}  \cdot \frac{\xi^+_k \ln(kn)}{c(r_u)c(r_w)}, \mbox{if $\varepsilon = r_{min}$,} \\
   D(r_u)^{\alpha_u}  \cdot \frac{\xi^+_k}{\gamma c(r_u)c(r_w)}, \mbox{if $\varepsilon < r_{min}$.}
   \end{cases}  \nonumber \\ & \leq
   \begin{cases}\\
   \frac{(\xi^+_k)^{\alpha_u+1}2^{(k+1)(\alpha_u+2)} k^{(k+2)(\alpha_u+1)}}{(\xi^-_k)^{\alpha_u+2}\gamma}n^{\alpha_u+r_w-\gamma-k}, \mbox{if $\varepsilon > r_{min}$,}\\
   \frac{(\xi^+_k)^{\alpha_u+1} 2^{(\alpha_u+2)(k+1)}k^{(k+1)\alpha_u+2} }{(\xi^-_k)^{\alpha_u+2}}n^{\alpha_u+r_w-\gamma-k}\ln(kn), \mbox{if $\varepsilon = r_{min}$,} \\
   \frac{(\xi^+_k)^{\alpha_u+1}2^{(\alpha_u+2)(k+1)}k^{(k+1)\alpha_u+3}}{(\xi^-_k)^{\alpha_u+2}\gamma} n^{\alpha_u+r_w-\gamma-k}, \mbox{if $\varepsilon < r_{min}$.}
   \end{cases}
\label{enq:uklower}
\end{align}
\end{small}
We now consider the payoff
of a node $u$ from the set $V_{=0}$. We have:
\begin{small}
\begin{align}
    P_{u,V_{=0}}({\bf r}) =\sum_{v \in V_{=0}}
    \frac{d_M(u,v)^{-r_u}}{c(r_u)c(0)} 
    \le
    \sum_{v \in V\setminus \{u\}}\frac{d_M(u,v)^{-r_u}}{c(r_u)c(0)}=\frac{1}{c(0)}.\label{enq:r0}
\end{align}
\end{small}
It is easy to get:
\begin{small}
\begin{equation}
c(0)=n^k-1\geq n^k/2.\label{enq:c0}
\end{equation}
\end{small}
Thus, combining with bound \eqref{eqn:distsk},
we have the payoff of node $u$ gets from
$ V_{=0}$:
\begin{small}
\begin{align}
D(r_u)^{\alpha_u} P_{u,V_{=0}}({\bf r}) \leq  \left(\frac{2^{k+1}\xi^+_k k^{2+k}}{\xi^-_k} n\right)^{\alpha_u} \cdot \frac{n^k}{2} \leq \frac{2^{(k+1)\alpha_u+1}k^{(2+k)\alpha_u}(\xi^+_k)^{\alpha_u}}{(\xi^-_k)^{\alpha_u}} n^{\alpha_u-k}.\label{enq:uzerolower}
\end{align}
\end{small}
Combining the above bounds in Eq.\eqref{enq:uklower} and Eq.\eqref{enq:uzerolower} with Eq.\eqref{enq:payoffsum},
we see the payoff of $r_u<k$ in ${\bf r}$ is at most $O\left(n^{\alpha_u+r_w-\gamma-k}\ln(kn)\right)$.

\para{Payoff of $r_u > k$.}If node $u$ chooses $r_u > k$, let $\varepsilon = r_u-k \geq \gamma$. We have:
\begin{small}
\begin{align}
P_{u, V_{>0}}  = \sum_{v \in V_{>0}}
    \frac{d_M(u,v)^{-r_u-r_{v}}}{c(r_u)c(r_v)} \leq \frac{\sum_{v \in V_{>0}}d_M(u,v)^{-r_u-r_{v}}}{c(r_u)c(r_w)} 
    <\frac{\sum_{v \in V\setminus \{u\}}d_M(u,v)^{-r_u}}{c(r_u)c(r_w)} =
\frac{1}{c(r_w)}.\nonumber
\end{align}
\end{small}
Combining the above inequality with bounds ~\eqref{enq:distanceboundc}, \eqref{enq:distanceboundd}, \eqref{enq:pr2c} and \eqref{enq:pr2a} on $D(r_u)$, $c(r_u)$ and $c(r_w)$, respectively,
we have the payoff of node $u$ gets from
$ V_{>0}$:
\begin{small}
\begin{align}
D(r_u)^{\alpha_u} P_{u,V_{>0}} \leq &
\begin{cases}
   \left(\frac{\xi^+_kk}{2\gamma \xi^-_k}n^{1-\gamma}\right)^{\alpha_u}\cdot \frac{2^{k+1}k}{\xi^-_k n^{k-r_w}} =\frac{2^{k+1-\alpha_u}k^{\alpha_u+1}(\xi^+_k)^{\alpha_u}}{(\xi^-_k)^{1+\alpha_u}\gamma^{\alpha_u}}n^{\alpha_u+r_w-k-\alpha_u\gamma}
   & \mbox{if $\varepsilon < 1$,}\\
   \left(\frac{\xi^+_k}{\xi^-_k}\ln(2kn)\right)^{\alpha_u}\cdot \frac{2^{k+1}k}{\xi^-_k n^{k-r_w}} =\frac{2^{k+1}k(\xi^+_k)^{\alpha_u}}{(\xi^-_k)^{1+\alpha_u}}\frac{\ln^{\alpha_u}(2kn)}{n^{k-r_w}} & \mbox{if $\varepsilon \geq 1$.}
   \end{cases}
\label{enq:ubklower}
\end{align}
\end{small}
Combining the bounds~\eqref{enq:distanceboundc}, \eqref{enq:distanceboundd}, \eqref{enq:pr2c} and with bounds Eq.\eqref{enq:r0} and Eq.\eqref{enq:c0},
we have the payoff of node $u$ gets from
$ V_{=0}$:
\begin{small}
\begin{align}
D(r_u)^{\alpha_u} P_{u,V_{=0}} \leq &
\begin{cases}
   \left(\frac{\xi^+_kk}{2\gamma \xi^-_k}n^{1-\gamma}\right)^{\alpha_u}\cdot \frac{2}{n^k}= 2^{1-\alpha_u}\left(\frac{\xi^+_kk}{\xi^-_k\gamma}\right)^{\alpha_u} n^{\alpha_u-k-\alpha_u\gamma}
   & \mbox{if $\varepsilon < 1$,}\\
   \left(\frac{\xi^+_k}{\xi^-_k}\ln(2kn)\right)^{\alpha_u}\cdot \frac{2}{n^k} = 2\left(\frac{\xi^+_k}{\xi^-_k}\right)^{\alpha_u}\frac{\ln^{\alpha_u}(2kn)}{n^k} & \mbox{if $\varepsilon \geq 1$.}
   \end{cases}
\label{enq:ubkzero}
\end{align}
\end{small}
Combining the above bounds in Eq.\eqref{enq:ubklower} and Eq.\eqref{enq:ubkzero} with Eq.\eqref{enq:payoffsum},
we see the payoff of $r_u > k$ is at most $O\left(n^{\alpha_u+r_w-k-\alpha_u\gamma}\right)$.

\para{Payoff of $r_u = k$.}However, if the node $u$ chooses the strategy $r_u=k$, we have:
\begin{small}
\begin{align}
P(r_u) > p_u(v, r_u)p_w(u, r_w) > \frac{d(u,w)^{-k-r_w}}{c(k)c(r_w)} > \frac{\delta^{-2k}}{c(k)c(r_w)}. \nonumber
\end{align}
\end{small}
Combining the above inequality with the bounds \eqref{enq:distanceboundb}, \eqref{enq:pr2b} and \eqref{enq:pr2a} on  $D(r_u)$, $c(k)$ and $c(r_w)$, we get:
\begin{small}
\begin{align}
\pi(r_u = k, \mathbf{r}_{-u}) & = D(r_u)^{\alpha_u} P(r_u) > \left(\frac{\xi_k^- n}{2c(k)}\right)^{\alpha_u} \cdot \frac{\delta^{-2k}}{c(k)c(r_w)} =\frac{(\xi_k^-)^{\alpha_u}\delta^{-2k}n^{\alpha_u}}{2^{\alpha_u} c^{1+\alpha_u}(k)c(r_w)} \nonumber \\& \ge \frac{(\xi_k^-)^{\alpha_u-1}k2^{1+k-\alpha_u}\delta^{-2k}}{(\xi^+_k)^{1+\alpha_u}}\frac{n^{\alpha_u+r_w-k}}{\ln^{1+\alpha_u}(2kn)}.
\end{align}
\end{small}
We see that the payoff of $r_u=k$ is in strictly higher order in $n$ than the payoff of $r_u<k$ or $r_u > k$,
thus there exists $n_0\in \mathbb{N}$ (which may depend on $\delta$ but do not
	depend on $r_w$ since $n^{\alpha_u+r_w-k}$ is a common term), for all $n\ge n_0$,
	$r_u=k$ is the best response to $\bf r_{-u}$ for any $u \in N_{w,\delta}$.
\end{proof}

\section{Proof of Random Small World Equilibrium}\label{proof1}
\response*
\begin{proof}[for $B_u({\mathbf{r}_{-u}} \equiv s)= 0$ if $s=0$]
When other players choose strategy $\mathbf{r}_{-u} \equiv 0$, the
reciprocity of the player $u$ is constant:

\begin{small}
\begin{equation}
P_u(r_u,{\bf r}_{-u}\equiv 0)=\sum_{\forall v \neq u}{p_u(v, r_u)p_v(u, r_v=0)} = \sum_{\forall v \neq
u}\frac{p_u(v,r_u)}{|V|-1} = \frac{1}{|V|-1}.
\end{equation}
\end{small}

Thus, the payoff of the player $u$ is only determined by the link
distance $D(r_u)$.
Let $X(r_u)$ be the random variable denoting the grid distance from
    $u$'s long-range contact to $u$.
Then we have $D(r_u) = E[X(r_u)]$.
We want to show the following intuitive claim:

{\em Claim 1.} For any $r_u < r_u'$, $X(r_u)$ strictly
stochastically dominates
    $X(r_u')$, i.e., for all $1 \le \ell <n_D$, $\Pr(X(r_u) \le \ell)
    < \Pr(X(r_u') \le \ell)$.

Proof of the claim. Let $q(r_u, j)$ be the probability that
    $u$'s long-range contact is a particular node $v$ at grid distance
    $j$ from $u$.
By definition, $q(r_u,j) = j^{-r_u}/c(r_u)$.
Then we have
$\frac{q(r_u,j+1)}{q(r_u, j)}= \left(\frac{j+1}{j}\right)^{-r_u}$.
Thus $q(r_u,j)$ is non-increasing in $j$, and the decreasing ratio is faster
    when $r_u$ is larger.
Since we know that $\sum_{j=1}^n q(r_u,j)b_u(j) = 1$, it must be that
    $q(r_u,1) < q(r_u',1)$, $q(r_u,n_D) > q(r_u',n_D)$, and there exists
    a $\bar{j}$ such that for all $j\le \bar{j}$,
    $q(r_u,j) \le q(r_u',j)$, and for all $j > \bar{j}$,
    $q(r_u,j) > q(r_u', j)$.

By the definition of $X(r_u)$, we have $\Pr(X(r_u) \le \ell)
    = \sum_{j=1}^\ell q(r_u,j)b_u(j)$.
Thus, for any $1 \le \ell \le \bar{j}$,
    $\Pr(X(r_u) \le \ell) = \sum_{j=1}^\ell q(r_u,j)b_u(j)
        < \sum_{j=1}^\ell q(r_u',j)b_u(j) = \Pr(X(r_u') \le \ell)$.
For any $ \bar{j} < \ell < n_D$,
    $\Pr(X(r_u) \le \ell) = \sum_{j=1}^\ell q(r_u,j)b_u(j)
        = 1 - \sum_{j=\ell+1}^{n_D} q(r_u,j)b_u(j)
        < 1 - \sum_{j=\ell+1}^{n_D} q(r_u',j)b_u(j) = \Pr(X(r_u') \le \ell)$.
Therefore, we have the claim that $X(r_u)$ strictly stochastically dominates
    $X(r_u')$.

With this claim, we immediately have $E[X(r_u)] > E[X(r_u')]$. As a
consequence, $D(0) = E[X(0)] >  E[X(r_u')] = D(r_u')$ for any
$r_u'>0$. Therefore, $r_u=0$ is the player $u$'s unique best
response to $\mathbf{r}_{-u} \equiv 0$.
%
%
%
%
\end{proof}

\section{Proof of Theorem 3.3}\label{proof4}
\strongNE*

\begin{proof}
We actually prove a slightly stronger result: any node $u$ in any strategy profile $\bf r$ with $r_u\ne k$
    is strictly worse off than its payoff in the navigable equilibrium,
    when $n$ is large enough.
With the Lemma~\ref{lem:payoffbound}, we see that a player $u$ has the payoff at least $\Omega\left(\frac{n^{\alpha_u}}{\ln^{2+\alpha_u}(2kn)}\right)$ before derivation.
Suppose that a coalition $C$ deviates, and the new strategy profile
    is $\bf r$.
Then some node $u\in C$ must select a new $r_u\ne k$.
By Lemma~\ref{lem:loworder}, there is a constant $\kappa$ such that
    for all sufficiently large $n$,
    $\pi(r_u, \mathbf{r}_{-u})\le \kappa n^{\alpha_u-\min\{1,\alpha_u\}\gamma}$.
Thus we see that the payoff of $u$ before
    the deviation is in strictly higher order in $n$ than its payoff
    after the deviation.
Therefore, for all sufficiently large $n$, $u$ is strictly worse off, which
    means no coalition could make some member strictly better off while
    others not worse off.
Hence, navigable small-world network (${\bf r}\equiv k$) is a strong Nash
    equilibrium.
\end{proof}

\section{Proof of Theorem 3.4}\label{proof6}
\coalition*
\begin{proof}
Given a pair of grid neighbors $(u,v)$, if they both choose the strategy $k$,
we have:
\begin{small}
\begin{align}
P(r_u, {\bf r_{-u}}) > p_u(v, r_u)p_v(u, r_u) \ge \frac{d_M(u,w)^{-2k}}{c(k)^2} \geq \frac{1}{c(k)^2}. \nonumber
\end{align}
\end{small}
Combining the above inequality with the bounds \eqref{enq:distanceboundb} and \eqref{enq:pr2b}, we get:
\begin{small}
\begin{align}
\pi(r_u = k, \mathbf{r}_{-u}) = D(r_u)^{\alpha_u}P(r_u, {\bf r_{-u}}) > \left(\frac{\xi_k^-n}{2}\right)^{\alpha_u} \frac{1}{c^{2+\alpha_u}(k)}\ge \frac{(\xi_k^-)^{\alpha_u}}{2^{\alpha_u}(\xi^+_k)^{\alpha_u+2}}\frac{n^{\alpha_u}}{\ln^{\alpha_u+2}(2kn)}.
\end{align}
\end{small}
However, if node $u$ chooses $r_u \neq k$, by Lemma~\ref{lem:loworder}
we know that there is a constant $\kappa$ such that
for all sufficiently large $n$, $\pi(r_u, \mathbf{r}_{-u})\le \kappa n^{\alpha_u-\min\{1,\alpha_u\}\gamma}$.
We see that the payoff of $r_u=k$ is in strictly higher order in $n$ than its original payoff.
Notice that the proof for the increase of node $v$'s payoff is similar to that of node $u$, so both colluding nodes get strictly higher payoff. The
theorem is proved.
\end{proof}

\section{Proof of Theorem 3.6}\label{proof7}
\begin{fact}
(Chernoff bound). Let $X$ be a sum of $n$ independent random
variables $\{X_i\}$, with $E[X_i]$ = $\mu$; $X_i \in \{0,1\}$ for
all $i\leq n$. For any $0<\epsilon<1,$
\begin{displaymath}Pr[X\leq(1-\epsilon)\mu]\leq
e^{-\frac{\mu\epsilon^2}{2}}, Pr[X\geq(1+\epsilon)\mu]\leq
e^{-\frac{\epsilon^2}{2+\epsilon}\mu}.
\end{displaymath}
\end{fact}

Based on the Chernoff bound, we have the following lemma. Let
$Y_u(j,s)$ be the number of players with grid distance $j$ to $u$
and a strategy of $s$.
\begin{lemma} \label{lem:high_prob}
In the $k$-dimensional DRB game $(k>1)$, for any $\eta > 0$, if each player chooses a
    strategy $s$ independently
    with probability $p_s \geq \eta$
	from a finite strategy set $S\subseteq \Sigma $,
	then for all $n\ge |S|$, with
	probability $1-1/n$, the following property holds:
\begin{displaymath}Y_u(j,s) > \frac{\eta b_u(j)}{2}, \forall u \in V,  \forall s \in S,
    \forall j \in \mathbb{N} \cap \left[\rho \left(\frac{\ln n}{\eta}\right)^{\frac{1}{k-1}}, \frac{n}{2}\right],
    \end{displaymath} where $\rho = \left(\frac{24+8k}{
\xi^-_k }\right)^{\frac{1}{k-1}}$ is a constant.
\end{lemma}

\begin{proof}
Since individual players choose strategy of $s$ independently with
probability $p_s$, $E[Y_u(j,s)] = p_s b_u(j)\geq \eta b_u(j)$. Based
on the Chernoff bound, we have:
\begin{small}
\begin{align}
P\left(Y_u(j)\leq(1-\epsilon)E[Y_u(j,s)]\right)\leq
\exp\left({-\frac{\epsilon^2 E[Y_u(j,s)]}{2}}\right)
\leq \exp\left({-\frac{\epsilon^2 \eta b_u(j)}{2}}\right). \nonumber
\end{align}
\end{small}
Note that $b_u(j) \geq \xi^-_k j^{k-1}$ for $0<j\leq \lfloor n
\rfloor/2$.
Let $m=|S|$.
Let $\varrho = \left(\frac{(16+8k)\ln n + 8\ln m}{\eta
\xi^-_k }\right)^{\frac{1}{k-1}}$ For $\varrho \leq j \leq \lfloor n
\rfloor/2$, we have:

\begin{small}
\begin{equation}
P\left(Y_u(j)\leq\frac{\eta b_j(u)}{2}\right)\leq \frac{1}{ m
n^{k+2}}. \nonumber
\end{equation}
\end{small}

Since there are $n^k$ players in the $k$ dimensional grid, by
    union bound, we have
    $\forall u, \forall s$, for any $\varrho \leq j \leq \lfloor n
\rfloor/2$,
\begin{small}
\begin{equation}
P\left(Y_u(j) > \frac{\eta b_j(u)}{2}\right)\geq 1-\frac{1}{n},\nonumber
\end{equation}
\end{small}

As $m$ is a constant, we can rewrite $\varrho$ as:
\begin{small}
\begin{equation}
\varrho= \left(\frac{(16+8k)\ln n + 8\ln m}{\eta \xi^-_k
}\right)^{\frac{1}{k-1}} \le \left(\frac{24+8k}{ \xi^-_k
}\right)^{\frac{1}{k-1}}\left(\frac{\ln
n}{\eta}\right)^{\frac{1}{k-1}}, \nonumber
\end{equation}
\end{small}
holds for all $n\geq m$.
\end{proof}

\perturbNE*
\begin{proof}
Given a deviation probability $p_u \le 1- n^{-\varepsilon}$ for node $u$, we know that node
	$u$ still uses the original strategy $k$ with a probability of
	$1-p_u \geq n^{-\varepsilon}$. By Lemma~\ref{lem:high_prob} we know that, with
probability $1-\frac{1}{n}$, the following property holds:
\begin{small}
\begin{equation}Y_u(j,k)
> \frac{n^{-\varepsilon}b_u(j)}{2}, \forall u \in V,
    \forall j \in \mathbb{N} \cap \left[\rho\left(\frac{\ln n}{n^{-\varepsilon}}\right)^{\frac{1}{k-1}}, \frac{n}{2}\right].
    \label{enq:density}
\end{equation}
\end{small}
When the above property holds, we fix any node $u$ and examine its payoff. In the case of $r_u=k$,
the reciprocity that $u$ gets from those still choosing strategy of
$k$ is:
\begin{small}
\begin{align}
P_{u, V_{=k}}({\bf r}) & \ge
\frac{n^{-\varepsilon}}{2}\frac{\sum_{j=\rho\left(\frac{\ln
n}{n^{-\varepsilon}}\right)^{\frac{1}{k-1}}}^{n/2}b_u(j)\cdot
j^{-2k}}{c^2(k)}  
\ge
\frac{\xi_k^-n^{-\varepsilon}}{2}\frac{\sum_{j=\rho\left(\frac{\ln
n}{n^{-\varepsilon}}\right)^{\frac{1}{k-1}}}^{n/2}
j^{-k-1}}{c^2(k)}  \nonumber\\
& \ge \frac{\xi^-_k n^{-\varepsilon}}{2}\frac{
\rho^{\frac{-(k+1)}{k-1}}\left(\frac{\ln
n}{n^{-\varepsilon}}\right)^{\frac{-(k+1)}{k-1}}}{c^2(k)} 
\ge \frac{\xi^-_k n^{-4\varepsilon}}{2c^2(k)\rho^3\ln^3{n}}. \nonumber
\end{align}
\end{small}
The last inequality holds as $k \ge 2$.

Combing with the above bound with bounds \eqref{enq:distanceboundb} and \eqref{enq:pr2b}, we get:
\begin{small}
\begin{align}
\pi_u(r_u=k,\mathbf{r}) & \geq D(r_u)^{\alpha_u} P_{u,V_{=k}}({\bf r}) \geq \left(\frac{\xi_k^- n}{2c(k)}\right)^{\alpha_u} \cdot \frac{\xi^-_k n^{-4\varepsilon}}{2c^2(k)\rho^3\ln^3{n}} \nonumber \\ & \geq \frac{(\xi^-_k)^{\alpha_u+1}}{2^{\alpha_u+1}(\xi^+_k)^{\alpha_u+2}\rho^3}\frac{n^{\alpha_u-4\varepsilon}}{ln^{2+\alpha_u}(2kn)\cdot ln^3(n)}
 \geq \frac{(\xi^-_k)^{\alpha_u+1}}{2^{\alpha_u+1}(\xi^+_k)^{\alpha_u+2}\rho^3}\frac{n^{\alpha_u-4\varepsilon}}{ln^{5+\alpha_u}(2kn)}.
\label{enq:prk}
\end{align}
\end{small}
By Lemma~\ref{lem:loworder}, there is a constant $\kappa$ such that
    for all sufficiently large $n$,
   $\pi(r_u\neq k, \mathbf{r}_{-u})\le \kappa n^{\alpha_u-\min\{1,\alpha_u\}\gamma}$.
Comparing with Eq.~\eqref{enq:prk},
    since $\alpha_{min} \leq \alpha_u$ for any node $u$ and $\varepsilon < \min\{1,\alpha_{min}\}\gamma/4$, the payoff of $u$
with strategy $r_u=k$ is in strictly higher order in $n$ than its
payoff after the deviation. Therefore, when the property Eq.~\eqref{enq:density} holds,
for all sufficiently large $n$,
$u$ get strictly better payoff than any other strategy choice by choosing $r_u=k$ after the deviation.

Therefore, when the property Eq.~\eqref{enq:density} holds,
the perturbed strategy profile ${\bf r}'$ moves back to the navigable
    small world (${\bf r}\equiv k$) in one synchronous step, as every
player $u$ moves from its current strategy to its best response $r_u=k$.
Also, it is clear that the property Eq.~\eqref{enq:density} consistently holds
as any player takes one asynchronous step. This is because the asynchronous move only increases the number of
 nodes choosing the strategy of $k$, so the best response of every player is always $k$
	after every asynchronous step. Thus,
the perturbed strategy profile moves back to the navigable
    small world as soon as every node takes at least one
asynchronous step. Notice that the property Eq.~\eqref{enq:density} holds with a probability of $1-1/n$,
so the theorem is proved.
\end{proof}

\section{Proof of Theorem 3.7}\label{proof8}
\perturbzero*
\begin{proof}
Fix any node $u\in V$.
Let $\bf r$ be the strategy profile after perturbation.
We partition nodes in $V\setminus \{u\}$ into
    sets $V_s, s \in S \cup \{0\}$, where $V_{s}
= \{v\in V \setminus \{u\} \mid r_v = s\}$.
Let $P_{u,V_s}({\bf r})$ be the
reciprocity $u$ obtained from subset $V_s$.
Then we have
\begin{small}
\begin{equation}
\pi(r_u, \mathbf{r}_{-u}) = D(r_u)\cdot \sum_{s\in S\cup \{0\}}{P_{u,V_s}({\bf
r})}. \label{eq:partition}
\end{equation}
\end{small}

For any node $u$ and any given $s\in S \cup \{0\}$, we now compare the payoff it gets from $V_{s}$
when using $r_u=k$ and $r_u=s'\neq k$, respectively.
\begin{small}
\begin{align}
\left(\frac{D(r_u=s')}{D(r_u=k)}\right)^{\alpha_u} \cdot \frac{P_{u,V_s}({\bf
r})}{P_{u,V_s}({\bf
r})} = \left(\frac{D(r_u=s')}{D(r_u=k)}\right)^{\alpha_u}\cdot \frac{\sum_{v \in V_{s}}\frac{d(u,v)^{-s'-s}}{c(s')c(s)}}{\sum_{v \in V_{s}}\frac{d(u,v)^{-k-s}}{c(k)c(s)}}.\nonumber
\end{align}
\end{small}

For a given node $u$ and a subset of nodes $\Gamma$, let define $d_{\min,\Gamma}$ and $d_{\max,\Gamma}$
be the minimum and maximum grid distances between node $u$ and any node $v \in \Gamma$, respectively. In other words,
$d_{\min,\Gamma} \leq d_M(u,v) \leq d_{\max,\Gamma}, \forall v \in \Gamma$.
With this definition, for any $v \in V_s$, we have:
\begin{small}
\begin{align}
\frac{\frac{d(u,v)^{-s'-s}}{c(s')c(s)}}{\frac{d(u,v)^{-k-s}}{c(k)c(s)}}
= \frac{c(k)}{c(s')}d(u,v)^{k-s'}\leq \begin{cases}
   \frac{c(k)}{c(s')}d_{\max,V_s}^{k-s'}
   &\mbox{if $s' < k$,}\\
   \frac{c(k)}{c(s')}d_{\min,V_s}^{k-s'}  &\mbox{if $s' > k$.}
   \end{cases}.\nonumber
\end{align}
\end{small}
Combing the above inequality, we have:
\begin{small}
\begin{equation}
\left(\frac{D(r_u=s')}{D(r_u=k)}\right)^{\alpha_u} \cdot \frac{P_{u,V_s}({\bf
r})}{P_{u,V_s}({\bf
r})} \leq \begin{cases} \left(\frac{D(r_u=s')}{D(r_u=k)}\right)^{\alpha_u}\cdot\frac{c(k)}{c(s')}d_{\max,V_s}^{k-s'}
   &\mbox{if $s' < k$,}\\
    \left(\frac{D(r_u=s')}{D(r_u=k)}\right)^{\alpha_u}\cdot\frac{c(k)}{c(s')}d_{\min,V_s}^{k-s'} &\mbox{if $s' > k$.}
   \end{cases}.\label{eq:compare_fraction}
\end{equation}
\end{small}
We first show that $\pi(r_u= k , \mathbf{r}_{-u}) > \pi(r_u=s' , \mathbf{r}_{-u})$ when $s'>k$.
In the case of $s'>k$, as $d_{\min,V_s} \geq 1$, combining the above inequality with
the bounds \eqref{enq:distanceboundc},  \eqref{enq:distanceboundb},  \eqref{enq:pr2c} and \eqref{enq:pr2b} on $D(s')$, $D(k)$, $c(s')$ and $c(k)$, we get:
\begin{small}
\begin{align}
\left(\frac{D(r_u=s')}{D(r_u=k)}\right)^{\alpha_u} \cdot \frac{P_{u,V_s}({\bf
r})}{P_{u,V_s}({\bf
r})} = O\left(\left(\frac{n^{1-\gamma}}{n}\right)^{\alpha_u}\cdot c(k)^{\alpha_u+1}\right) = O\left(\frac{\ln^{\alpha_u+1}(2kn)}{n^{\alpha_u\gamma}}\right).
\label{big_bound}
\end{align}
\end{small}
Therefore, we can find a constant $\sigma$ such that:
\begin{small}
\begin{align}
& \pi(r_u= k , \mathbf{r}_{-u})  -  \pi(r_u=s' , \mathbf{r}_{-u})  = \sum_{s\in S\cup \{0\}}\left[D(r_u=k)^{\alpha_u}P_{u,V_s}({\bf
r})- D(r_u = s')^{\alpha_u}P_{u,V_s}({\bf
r})\right]  \nonumber \\
& = \sum_{s\in S\cup \{0\}} D(r_u=k)^{\alpha_u}P_{u,V_s}({\bf
r})\left(1 - \frac{D(r_u=s')^{\alpha_u}P_{u,V_s}({\bf
r})}{D(r_u=k)^{\alpha_u}P_{u,V_s}({\bf
r})}\right)  \nonumber \\
& \geq \sum_{s\in S\cup \{0\}}D(r_u=k)^{\alpha_u}P_{u,V_s}({\bf
r})\left(1-\frac{\ln^{\alpha_u+1}(2kn)}{n^{\alpha_u\gamma}}\right) > 0,
\label{eq:case1}
\end{align}
\end{small}
for sufficiently large $n$.

We next show that $\pi(r_u= k , \mathbf{r}_{-u}) > \pi(r_u=s' , \mathbf{r}_{-u})$ when $s' < k$.
Note here we require the constant $\varepsilon < \gamma/2$ in the theorem. We first find a distance threshold
to partition nodes into nodes nearby to $u$ and nodes far away from $u$. We want to prove that
$\pi(r_u= k , \mathbf{r}_{-u}) - \pi(r_u=s' , \mathbf{r}_{-u})$ is dominated by
the nearby nodes.

In the case of $s' < k$, we can find a constant $\nu = 1-\frac{\gamma-2\varepsilon}{2k}$ such that,
for any $s \in S$, the set $V_s$ can be partitioned into two subsets: (i) $V_s^- = \{v\in V\mid r_v=s \wedge d_M(u,v) \leq n^\nu\}$,
and (ii) $V_s^+ = \{v\in V\mid r_v=s \wedge d_M(u,v) > n^\nu \}$.
Notice that $d_{\max, V_s^-}$ is at most $ n^\nu$.
Combining the above inequality Eq.~\eqref{eq:compare_fraction} with
the bounds \eqref{enq:distancebounda},  \eqref{enq:distanceboundb},  \eqref{enq:pr2a} and \eqref{enq:pr2b} on $D(s')$, $D(k)$, $c(s')$ and $c(k)$,
we get:
\begin{align}
\left(\frac{D(r_u=s')}{D(r_u=k)}\right)^{\alpha_u} \cdot \frac{P_{u,V_s^-}({\bf
r})}{P_{u,V_s^-}({\bf
r})} = O\left(\frac{c(k)^{\alpha_u+1}}{n^{(k-s')(1-\nu)}}\right) = O\left(\frac{\ln^{\alpha_u+1}(2kn)}{n^{(1-\nu)\gamma}}\right),
\label{small_bound}
\end{align}
where $\nu < 1.$

Notice that $P_{u,V_s} = P_{u,V_s^-}+P_{u,V_s^+}$. Based on the bound
	in Eq.~\eqref{small_bound},
we can find a constant $\sigma'$ such that:
\begin{small}
\begin{align}
& \pi(r_u= k , \mathbf{r}_{-u})  -  \pi(r_u=s' , \mathbf{r}_{-u})  \nonumber \\
& \geq \sum_{s\in S\cup \{0\}}D(r_u=k)^{\alpha_u}P_{u,V_s^-}({\bf
r})- \sum_{s\in S\cup \{0\}}D(r_u = s')^{\alpha_u}P_{u,V_s}({\bf
r})  \nonumber \\
& \geq \sum_{s\in S\cup \{0\}}D(r_u=k)^{\alpha_u}P_{u,V_s^-}({\bf
r})\left(1 - \frac{D(r_u=s')^{\alpha_u}P_{u,V_s^-}({\bf
r})}{D(r_u=k)^{\alpha_u}P_{u,V_s^-}({\bf
r})}\right) - \sum_{s\in S\cup \{0\}}D(r_u = s')^{\alpha_u}P_{u,V_s^+}({\bf
r})  \nonumber \\
& \geq \sum_{s\in S\cup \{0\}}D(r_u=k)^{\alpha_u}P_{u,V_s^-}({\bf
r})\left(1 - \frac{\sigma'\ln^{\alpha_u+1}(2kn)}{n^{(1-\nu)\gamma}}\right) - \sum_{s\in S\cup \{0\}}D(r_u = s')^{\alpha_u}P_{u,V_s^+}({\bf
r})  \nonumber \\
& \geq \sum_{s\in S\cup \{0\}}\frac{D(r_u=k)^{\alpha_u}P_{u,V_s^-}({\bf
r})}{2}
- \sum_{s\in S\cup \{0\}}D(r_u = s')^{\alpha_u}P_{u,V_s^+}({\bf
r})
\label{eq:minus}
\end{align}
\end{small}
for sufficiently large $n$.

We now give the lower bound of the first term $D(r_u=k)^{\alpha_u}P_{u,V_s^-}({\bf r})$.
Let $U_{j} = \{v \mid d_M(u,v)=j\wedge r_v >0\}$.
By Lemma~\ref{lem:high_prob}, with probability $1-1/n$,
\begin{small}
\begin{equation}|U_{j}|
> \frac{\eta b_u(j)}{2}, \forall u \in V,
    \forall j \in \mathbb{N} \cap \left[\rho\left(\frac{\ln n}{\eta}\right)^{\frac{1}{k-1}}, \frac{n}{2}\right].
    \label{enq:nonzero_density}
\end{equation}
\end{small}
For $j=\left\lceil \rho\left(\frac{\ln n}{\eta}\right)^{\frac{1}{k-1}} \right\rceil$, we have:
\begin{small}
\begin{align*}
P_{u, U_{j}}({\bf r})
& = \sum_{v\in U_j} p_u(v, r_u) \cdot p_v(u, r_v) = \sum_{v\in U_j} \frac{j^{-k}}{c(k)}\cdot \frac{j^{-r_v}}{c(r_v)}\ge \sum_{v\in U_j} \frac{j^{-k}}{c(k)}\cdot \frac{j^{-\beta}}{c(\gamma)} \ge
\frac{\eta \cdot b_u(j) \cdot j^{-k-\beta}}{2c(k)c(\gamma)} \\
& \ge \frac{\eta \cdot \xi^-_k j^{k-1}  \cdot j^{-k-\beta}}{2c(k)c(\gamma)}
\ge
\frac{\eta\xi^-_k \rho^{\frac{-(\beta+1)}{k-1}}\left(\frac{\ln
n}{\eta}\right)^{\frac{-(\beta+1)}{k-1}}}{2c(k)c(\gamma)}.
\end{align*}
\end{small}
We now fix $\eta = 1/n^\frac{(k-1)\varepsilon}{k+\beta} (0
<\varepsilon < \gamma/2)$, and have:
\begin{small}
\begin{equation}
P_{u, U_{j}}({\bf r}) \ge  \frac{\xi^-_k
}{2\rho^{\frac{\beta+1}{k-1}}c(k)c(\gamma)} \cdot \frac{1}{(\ln
n)^{\frac{(\beta+1)}{k-1}} n^{\varepsilon}}.
\end{equation}
\end{small}

Note that $ U_{j} \subseteq \cup_{s\in S\setminus\{0\}} V_s^-$,
since $j = \left\lceil \rho\left(\frac{\ln
n}{\eta}\right)^{\frac{1}{k-1}} \right\rceil =
\left\lceil \rho \ln
^\frac{1}{k-1}n \cdot n^{\frac{\varepsilon}{k+\beta}}\right\rceil
< n^{\frac{k-\gamma/2+\varepsilon}{k}} = n^\nu$ for sufficiently large $n$.
Combining with the bounds \eqref{enq:distanceboundb}, \eqref{enq:pr2a} and \eqref{enq:pr2b} on $D(k)$, $c(\gamma)$ and $c(k)$, we get:
\begin{small}
\begin{align}
D(r_u=k)^{\alpha_u}\sum_{s\in S\setminus \{0\}}P_{u,V_s^-} &  \geq D(r_u=k)^{\alpha_u} P_{u,U_{j}}(r_u=k, {\bf r}_{-u}) \nonumber \\ &= \Omega\left(\frac{n^{\alpha_u}}{\ln^{\alpha_u}(2kn)} \cdot \frac{1}{\ln(2kn)\cdot n^{k-\gamma}} \cdot \frac{1}{(\ln
n)^{\frac{(\beta+1)}{k-1}} n^{\varepsilon}} \right)= \Omega\left(
\frac{n^{\alpha_u-k-\varepsilon+\gamma}}{\ln^a(2kn)}\right), \label{eq:short_payoff}
\end{align}
\end{small}
where $a= \alpha_u + 1 +\frac{\beta+1}{k-1}$
is constant.

We next give the upper bound of the second term $D(r_u = s')^{\alpha_u}P_{u,V_s^+}$.
Notice that $d_M(u,v) > n^\nu$ for any $v$ in $V_s^+$, so for any $s$, we have:
\begin{small}
\begin{align}
& P_{u,V_s^+}({\bf
r}) = \sum_{v \in V_s^+}\frac{d_M(u,v)^{-s'-s}}{c(s')c(s)} \leq \sum_{v \in V_s^+} \frac{n^{-\nu(s'+s)}}{c(s')c(s)}.
\label{enq:farreciprocity}
\end{align}
\end{small}
In the case of $s<k$, combining the above
inequality with the \chgins{bound \eqref{enq:pr2a} on $c(s')$ and $c(s)$}, we get:
\begin{small}
\begin{align}
P_{u,V_s^+}({\bf
r}) \leq \sum_{v \in V_s^+} \frac{2^{2k+2}k^2}{(\xi_k^-)^2}n^{(s+s')(1-\nu)-2k}
 \leq \sum_{v \in V_s^+}\frac{2^{2k+2}k^2}{(\xi_k^-)^2}n^{2k(1-\nu)-2k}= | V_s^+|\frac{2^{2k+2}k^2}{(\xi_k^-)^2}n^{-2k\nu}.
\label{enq:reciprocity_small}
\end{align}
\end{small}
In the other case of $s\geq k$, combining the
inequality Eq.~\eqref{enq:farreciprocity} with the bounds \eqref{enq:pr2a}, \eqref{enq:pr2c} on $c(s')$ and $c(s)$, respectively, we get:
\begin{small}
\begin{align}
 P_{u,V_s^+}({\bf
r})& \leq  \sum_{v \in V_s^+} \frac{n^{-\nu(s'+k)}}{c(s')\xi_k^-} \leq
\sum_{v \in V_s^+} \frac{2^{k+1}k}{(\xi_k^-)^2}n^{(1-\nu)s'-(1+\nu)k} \leq \sum_{v \in V_s^+}\frac{2^{k+1}k}{(\xi_k^-)^2}n^{(1-\nu)k-(1+\nu)k}\nonumber \\
&= | V_s^+|\frac{2^{k+1}k}{(\xi_k^-)^2}n^{-2k\nu}.
\label{enq:reciprocity_large}
\end{align}
\end{small}

Combining the above
inequalities Eq.~\eqref{enq:reciprocity_small} and Eq.~\eqref{enq:reciprocity_large} with the bound \eqref{enq:distanceboundb} on distance $D(s')$,
we know that for any $s$:
\begin{small}
\begin{align}
& D(r_u = s')^{\alpha_u}P_{u,V_s^+}
 = O\left(|V_s^+|n^{\alpha_u-2k\nu}\right).
\label{eq:long_payoff}
\end{align}
\end{small}

We are now ready to combine the above bounds and show that $\pi(r_u= k , \mathbf{r}_{-u}) > \pi(r_u=s' , \mathbf{r}_{-u})$ when $s' < k$.
More specifically, combining the inequalities in Eq.~\eqref{eq:minus}, Eq.~\eqref{eq:short_payoff} and
Eq.~\eqref{eq:long_payoff}, we get:
\begin{small}
\begin{align}
& \pi(r_u= k , \mathbf{r}_{-u})  -  \pi(r_u=s' , \mathbf{r}_{-u}) \nonumber \\
&  \geq  \sum_{s\in S\setminus \{0\}}\frac{D(r_u=k)^{\alpha_u}P_{u,V_s^-}}{2}-  \sum_{s\in S\cup \{0\}} D(r_u = s')^{\alpha_u}P_{u,V_s^+},\nonumber \\
& \geq \frac{\rho n^{\alpha_u-k+\gamma-\varepsilon}}{2\ln^a(2kn)} - \rho' |\cup_{s\in S \cup \{0\}} V_s^+| \cdot n^{\alpha_u-2k\nu}  \geq \frac{\rho n^{\alpha_u-k+\gamma-\varepsilon}}{2\ln^a(2kn)} - \rho' n^{k} \cdot n^{\alpha_u-2k + \gamma - 2\varepsilon} \nonumber \\
& \geq \frac{\rho n^{\alpha_u-k+\gamma-\varepsilon}}{2\ln^a(2kn)} - \rho' n^{\alpha_u-k + \gamma - 2\varepsilon},
\label{eq:finalminus}
\end{align}
\end{small}
where $\sigma, \rho, \rho', a$ are all constants.

As $0 < \varepsilon < \gamma/2$, the first term in Eq.~\eqref{eq:finalminus} is in strictly higher
    order in $n$ than the second term in Eq.~\eqref{eq:finalminus}, we know
    that for sufficiently large $n$,
    $\pi(r_u= k , \mathbf{r}_{-u}) > \pi(r_u=s' , \mathbf{r}_{-u})$ for any $s' < k$.

Therefore, when the property in Eq.~\eqref{enq:nonzero_density} holds,
the perturbed strategy profile ${\bf r}$ moves to the navigable
    small world (${\bf r}'\equiv k$) in one synchronous step, as every
player $u$ moves from its current strategy to its best response $r'_u=k$.
Also, it is clear that the property Eq.~\eqref{enq:nonzero_density} consistently holds
as any player takes one asynchronous step. This is because the asynchronous move only increases the number of
 nodes choosing a non-zero strategy, so the best response of every player is always $k$
	after every asynchronous step. Thus,
the perturbed strategy profile moves to the navigable
    small world as soon as every node takes at least one
asynchronous step. Notice that the property in Eq.~\eqref{enq:nonzero_density} holds with a probability of $1-1/n$,
so the theorem is proved.
\end{proof}

\section{Proof of Theorem 4.1}\label{proof9}
\optimal*
\begin{proof}
Given the strategy profile $\bf r$, we partition the nodes $V$ into three sets:
$V_{< k} = \{v\in V \mid r_v < k\}$, $V_{> k} = \{v\in V \mid r_v > k\}$,
$V_{ = k} = \{v\in V \mid r_v = k\}$. 
So we have:
\begin{small}
\begin{equation}
\pi_u(\mathbf{r}) = D(r_u)^\alpha\left( P_{u,V_{<k}}({\bf r}) +
    P_{u,V_{>k}}({\bf r}) + P_{u,V_{=k}}({\bf r}) \right).
\end{equation}
\end{small}

For any node $u\in V_{= k}$, we have:
\begin{small}
\begin{align}
P_{u, V_{ < k}}({\bf r}) &  =  \sum_{v \in V_{ < k}}
    \frac{d_M(u,v)^{-r_u-r_v}}{c(r_u)c(r_v)}
    \le \sum_{v \in V_{ < k}}\frac{d_M(u,v)^{-r_u}}{c(r_u)c(k-\gamma)} \leq \frac{\sum_{\forall v \ne u}
d_M(u,v)^{-r_u}}{c(r_u)c(k-\gamma)} = \frac{1}{c(k-\gamma)}. \label{enq:prr} \nonumber
\end{align}
\end{small}
Combining the above inequality with bounds \eqref{enq:distanceboundb}, \eqref{enq:pr2b} and \eqref{enq:pr2a} on $D(k)$, $c(k)$ and $c(k-\gamma)$,
we get the upper bound on the payoff obtained from the set $V_{<k}$:
\begin{small}
\begin{equation}
D(r_u=k)^{\alpha} P_{u,V_{<k}}({\bf r}) \leq (\frac{2\xi_k^+ n}{\xi^-_k\ln n} )^{\alpha} \cdot  \frac{2^{k+1}k}{\xi^-_k n^{\gamma}} \leq
\frac{(\xi^+_k)^{\alpha} 2^{k+1+\alpha}k}{(\xi^-_k)^{\alpha+1}}\frac{n^{\alpha-\gamma}}{\ln^\alpha n}.
\label{enq:uklower<k}
\end{equation}
\end{small}

For the set $V_{>k}$, we have:
\begin{small}
\begin{equation}
\begin{aligned}
P_{u, V_{>k}}({\bf r})
& =  \sum_{v \in V_{ > k}}
    \frac{d_M(u,v)^{-r_u-r_v}}{c(r_u)c(r_v)}  \leq \frac{\sum_{j=1}^{n_D}{b_u(j)\cdot j^{-r_u} \cdot
j^{-k-\gamma}}}{\xi^-_k c(r_u)} =
\frac{\xi^+_k\sum_{j=1}^{n_D}j^{-1-r_u-\gamma}}{\xi^-_kc(r_u)}\\
& \leq
\frac{\xi^+_k(1+\int_{1}^{n_D}x^{-1-r_u-\gamma}dx)}{\xi^-_kc(r_u)} \leq
\frac{\xi^+_k(1+r_u+\gamma)}{\xi^-_k(r_u+\gamma) c(r_u)} \leq
\frac{\xi^+_k(k+1)}{\xi^-_k\gamma c(r_u)}. \label{enq:prbr} \nonumber
\end{aligned}
\end{equation}
\end{small}

Combining the above inequality with the bounds \eqref{enq:distanceboundb}, \eqref{enq:pr2b} on $D(k)$, $c(k)$,
we get the upper bound on the payoff obtained from the set $V_{>k}$:
\begin{small}
\begin{equation}
D(r_u=k)^{\alpha}P_{u,V_{>k}}({\bf r}) \leq  (\frac{2\xi_k^+ n}{\xi^-_k\ln n} )^{\alpha} \cdot  \frac{2\xi^+_k(k+1)}{(\xi^-_k)^2\gamma \ln n} \leq
\frac{2^{\alpha+2}(\xi^+_k)^{\alpha+1}k}{\gamma(\xi^-_k)^{\alpha+2}}\frac{n^\alpha}{\ln^{\alpha+1} n}.
\label{enq:uklowersk}
\end{equation}
\end{small}

For the set $V_{=k}$, by Lemma~\ref{lem:payoffbound} we know that:
\begin{small}
\begin{align}
D(r_u=k)P_{u,V_{=k}\setminus \{u\}}({\bf r}) \leq \pi(r_u = k, \mathbf{r}_{-u} \equiv k) \leq \frac{2^{\alpha+3}(\xi_k^+)^{\alpha+1}}{(\xi_k^-)^{2+\alpha}}\cdot\frac{n^{\alpha}}{\ln^{\alpha+2}n}
\label{enq:uklowerlk}
\end{align}
\end{small}

Then, for any node $u\in V_{=k}$ and sufficiently large $n$, we have
\begin{small}
\begin{align}
&\pi_u(\mathbf{r}) = D(r_u)^\alpha\left( P_{u,V_{<k}}({\bf r}) +
    P_{u,V_{>k}}({\bf r}) + P_{u,V_{=k}\setminus \{u\}}({\bf r}) \right) \nonumber \\
& \leq \frac{(\xi^+_k)^{\alpha} 2^{k+1+\alpha}k}{(\xi^-_k)^{\alpha+1}}\frac{n^{\alpha-\gamma}}{\ln^\alpha n} +
\frac{2^{\alpha+2}(\xi^+_k)^{\alpha+1}k}{\gamma(\xi^-_k)^{\alpha+2}}\frac{n^\alpha}{\ln^{\alpha+1} n}
+  \frac{2^{\alpha+3}(\xi_k^+)^{\alpha+1}}{(\xi_k^-)^{2+\alpha}}\cdot\frac{n^{\alpha}}{\ln^{\alpha+2}n} \nonumber \\ &
 < \frac{2^{\alpha+3}(\xi^+_k)^{\alpha+1}k}{\gamma(\xi^-_k)^{\alpha+2}}\frac{n^\alpha}{\ln^{\alpha+1} n} .
\end{align}
\end{small}
By Lemma~\ref{lem:loworder}, for any node $u\notin V_{ = k}$, there is a constant $\kappa$ such that
for all sufficiently large $n$, $\pi(r_u, \mathbf{r}_{-u})\le \kappa n^{\alpha-\min\{1,\alpha\}\gamma}$.

So we have, for sufficiently large $n$, the social welfare of the profile is:
\begin{small}
\begin{align}
SW(\mathbf{r}) =  & \sum_{u\in V}\pi(r_u, \mathbf{r}_{-u}) < |V|\left(\frac{2^{\alpha+3}(\xi^+_k)^{\alpha+1}k}{\gamma(\xi^-_k)^{\alpha+2}}\frac{n^\alpha}{\ln^{\alpha+1} n} + n^{\alpha-\min\{1,\alpha\}\gamma}\right) \nonumber \\ &
< \frac{2^{\alpha+4}(\xi^+_k)^{\alpha+1}k}{\gamma(\xi^-_k)^{\alpha+2}}\frac{n^{\alpha+k}}{\ln^{\alpha+1} n}.
\label{enq:optimalupper}
\end{align}
\end{small}

The above inequality shows that the social welfare of the profile is at most $O\left(\frac{n^{\alpha+k}}{\ln^{\alpha+1} n}\right)$.

Next, we construct the profile $\bf r$ as follows: for any node with location $(i,j)$,
we set its strategy as $k$ if $j\mod 2 = 0$, otherwise,
we set its strategy as $k+\gamma$.

Notice that for any node $u$ with $r_u=k$, it has at least one neighbor $v$ with $r_v > k$.
We get:
\begin{small}
\begin{align}
P_{u,V} > p_u(v, r_u)p_v(u, r_v) > \frac{1}{c(k)c(k+\gamma)}.
\end{align}
\end{small}


Combining with bounds in \eqref{enq:distanceboundb}, \eqref{enq:pr2b} and  \eqref{enq:pr2c} on $D(k)$, $c(k)$ and $c(k+\gamma)$, for any node $u$ with $r_u=k$,
\begin{small}
\begin{align}
\pi_u(\mathbf{r}) >  D(r_u)^\alpha P_{u,V}({\bf r}) > \frac{(\xi^-_k)^\alpha\gamma}{2^\alpha(\xi^+_k)^{2+\alpha}(1+\gamma)}\frac{n^\alpha}{\ln^{\alpha+1}(2kn)} 
> \frac{(\xi^-_k)^\alpha\gamma}{2^{2\alpha+1}(\xi^+_k)^{2+\alpha}(1+\gamma)}\frac{n^\alpha}{\ln^{\alpha+1}(kn)}.
\end{align}
\end{small}

So we have, for sufficiently large $n$, the social welfare of the profile is:
\begin{small}
\begin{align}
SW(\mathbf{r}) =  \sum_{u\in V}\pi(r_u, \mathbf{r}_{-u}) > \frac{|V|}{2}  \frac{(\xi^-_k)^\alpha\gamma}{2^{2\alpha+1}(\xi^+_k)^{2+\alpha}(1+\gamma)}\frac{n^\alpha}{\ln^{\alpha+1}(kn)} 
 >
 \frac{(\xi^-_k)^\alpha\gamma}{2^{2\alpha+2}(\xi^+_k)^{2+\alpha}(1+\gamma)}\frac{n^{\alpha+k}}{\ln^{\alpha+1}(kn)}.
\label{enq:optimallower}
\end{align}
\end{small}

The above inequality shows that the optimal social welfare is at least $\Omega\left(\frac{n^{\alpha+k}}{\ln^{\alpha+1} n}\right)$.
Combining the results of Eq.\eqref{enq:optimalupper} and Eq.\eqref{enq:optimallower}, the theorem is proved.
\end{proof}

\section{Proof of Theorem 4.2}\label{proof10}
\worst*
\begin{proof}
According to Lemma~\ref{lem:payoffbound}, the payoff of each player in navigable NE ${\bf r} \equiv k$ is $\pi(r_u = k, \mathbf{r}_{-u} \equiv k)=\Theta\left(\frac{n^{\alpha}}{\ln^{\alpha+2}n}\right)$, so the social welfare of navigable NE is $\Theta(\frac{n^{\alpha+k}}{\ln^{\alpha+2}n})$.
Combining with Theorem~\ref{thm:optimal}, the price of stability (PoS)
is $\Theta({\ln n})$.


For random small world $\mathbf{r}_{-u}\equiv 0$, we have:
\begin{small}
\begin{align}
   P_u(r_u,{\bf r}_{-u}\equiv 0) & =\sum_{v \in V}
    \frac{d_M(u,v)^{-r_u}}{c(r_u)c(0)} =\frac{1}{c(0)}.\label{enq:zeroreciprocity}
\end{align}
\end{small}
It is easy to get:
\begin{small}
\begin{equation}
c(0)=n^k-1\geq n^k/2.\label{enq:zeroconstant}
\end{equation}
\end{small}
Thus, combining the above inequality with the distance bound~\eqref{enq:distancebounda},
we have the payoff of node $u$ gets from
$ V_{=0}$:
\begin{small}
\begin{align}
\pi(r_u,{\bf r}_{-u}\equiv 0) \leq \frac{2^{(k+1)\alpha+1}k^{\alpha(2+k)}(\xi^+_k)^\alpha}{(\xi^-_k)^\alpha} n^{\alpha-k}.\label{enq:uzerolowerbound}
\end{align}
\end{small}
According to the above inequality, it is easy to get that
	the social welfare of ${\bf r} \equiv 0$ is at most
$O(n^\alpha)$.
We now examine its lower bound. To do so, we first get the
lower bound on distance.
\begin{small}
\begin{align}
D(r_u=0)  \geq \frac{\sum_{j=1}^{n/2}b_u(j)\cdot j}{c(0)} \geq
\frac{\xi^-_k\int_{1}^{n/2}x^kdx}{c(0)} 
 \geq \frac{\xi^-_k (n/2-1)^{1+k}}{(1+k) c(0)}
    > \frac{\xi^-_k (n/4)^{1+k}}{(1+k)c(0)},
\end{align}
\end{small}
Combining the above inequality with Eq.\eqref{enq:zeroconstant} and Eq.~\eqref{enq:zeroreciprocity}, so we can get
\begin{small}
\begin{align}
\pi(r_u = k, \mathbf{r}_{-u}\equiv 0) > \frac{(\xi^-_k)^\alpha}{4^{\alpha(1+k)}2^{\alpha-1}k}n^{\alpha-k}.
\end{align}
\end{small}
Therefore, the social welfare of the random small-world network (${\bf r} \equiv 0$) is $\Theta(n^\alpha)$.
Combining with Theorem~\ref{thm:optimal}, the price of anarchy (PoA)
is $\Theta\left(\frac{n^k}{\ln^{\alpha+1}n}\right)$.
\end{proof}



\end{document}